\documentclass[12 pt]{article}
\usepackage{amsmath,amsthm,amssymb,amsfonts}
\usepackage[a4paper]{geometry}
\usepackage{a4wide}
\usepackage{parskip}
\usepackage{graphicx}
\usepackage{booktabs}
\usepackage{colortbl}

\usepackage{textcomp}
\usepackage{url}
\usepackage{hyperref}
\usepackage[numbers]{natbib}
\usepackage{color}
\usepackage{xcolor}
\usepackage{cleveref}
\usepackage{bm}
\usepackage{booktabs,tabularx,pbox}
\usepackage{tabulary}
\usepackage{array,longtable}
\newcolumntype{?}{!{\vrule width 1.5pt}}
\usepackage{multicol}
\usepackage{multirow}
\usepackage{boldline}
\usepackage{caption}
\usepackage{subcaption}
\usepackage[font=scriptsize,skip=0pt]{caption}
\usepackage[font=scriptsize,skip=0pt]{subcaption}
\usepackage{epsfig}
\usepackage{epstopdf}
\usepackage{array,longtable}
\usepackage{multicol}
\usepackage{supertabular}
\usepackage{lipsum}
\usepackage{float}

\title{\textbf{Exploratory data analysis for large-scale multiple testing problems and its application in gene expression studies}}
\author{Paramita Chakraborty\thanks{Corresponding author: 219A, Leconte College, 1523 Greene Street, Columbia, SC. \newline\indent\hspace{2.625 cm}  Email: chakrabp@stat.sc.edu (P. Chakraborty)} ,
Chong Ma,  John Grego and James Lynch.\vspace{0.2 in}\\ \textit{Department of Statistics, University of South Carolina, Columbia.} %\\ %\textit{$^2$Department of Biostatistics, Yale School of Public Health, Yale University.}
}
\date{}

\allowdisplaybreaks
\usepackage[none]{hyphenat}
\usepackage{epstopdf}
\usepackage{array,longtable}
\usepackage{multicol}
\usepackage{supertabular}
\usepackage{lipsum}
\usepackage{float}
\usepackage{bm}
\numberwithin{equation}{section}
\theoremstyle{plain}

\begin{document}
\maketitle

%%%%%%%%%%%%%%%%%%%
%%%%%%%%%%%%%%%%%%
\begin{abstract}
\noindent
In large scale multiple testing problems, a two-class empirical Bayes approach can be used to control the false discovery rate (Fdr) for the entire array of hypotheses under study.  A sample splitting step is incorporated to modify that approach where one part of the data is used for model fitting and the other part for detecting the significant cases by a screening technique featuring the empirical Bayes mode of Fdr control. Cases with high detection frequency across repeated random sample splits are considered true discoveries. A critical detection frequency is set to control the overall false discovery rate. The proposed method helps to balance out unwanted sources of variation and  addresses potential statistical overfitting of the core empirical model by cross-validation through resampling. Further, concurrent detection frequencies are used to provide visual tools to explore the inter-relationship between significant cases. The methodology is illustrated using a microarray data set, RNA-sequencing data set, and several simulation studies. A power analysis is presented to understand the efficiency of the proposed method.
\end{abstract}

%%%%%%%%%%%%%%%%%%%%%%%%%%%%

\section{Introduction}
\label{s_intro}
A multiple hypotheses scenario arises when a number of factors are believed to contribute to the difference between two populations. A typical example in biology is transcriptome analysis in which thousands of gene expression are examined to detect a handful that are expressed at significantly different levels between two groups of subjects with different biological outcomes. A common goal of such a study is to identify the regulator genes that influence the outcome of interest. Empirical Bayes methods can provide useful analysis tools for these problems.

The empirical Bayes model was considered by Robbins \cite{Robbins} to study iid observations $(X_i, \Theta_i)$, $i=1,2,\ldots,n$, where $\Theta_i$ is a latent variable. This model results in the mixture model where $X_1, X_2,\ldots$ are iid with density
\begin{equation}\label{eq:mix_density}
f(x)=\int f(x|\theta)\mu(d\theta).
\end{equation}
Here $f(x|\theta)$ is the conditional density of $X$ given $\Theta=\theta$ and $\mu$ is the marginal density of $\Theta$.

In addition to justifying shrinkage estimation \cite{Morris}, the method has been very useful in large scale inference \cite{Efron07, Efron08, Efron10}. The specialization of \eqref{eq:mix_density} to the two point mixture model has played an important role in the analysis of the two class data problem. In the model
\begin{equation}
\label{eq:main_model}
f(x)=p_0f_0(x)+p_1f_1(x),
 \end{equation}
$f_0$ is  the background density and $f_1$ is the signal or contamination density with $p_0+p_1=1$. This representation can be used for a generic multiple testing problem where the hypotheses under study are divided into the null class and the non-null class   \cite{Efron07} and one wishes to control the false discovery rate \cite{Benjamini} for the entire array of hypotheses under study. For example, a two point mixture model where $f_0$ is the \textit{Uniform} and $f_1$ is the \textit{Beta} distribution is used in \cite{Allison} to analyze p-values for gene expression data.

The methodology proposed in this paper is a modification of these ideas with the two-class empirical Bayes models for p-values in multiple testing problems. The modification considers the two-point  \textit{Uniform-Beta} mixture model \eqref{eq:main_model} and then adjusts the mixture component to accommodate better the extreme p-values. The main step incorporates repeated random sample splitting to use the ``modeling" split to fit this adjusted model. The fitted model then is used to identify the signal p-values in the ``screening" split. The reoccurrence of the signal p-values is finally used to ``identify" significant signals in the data. The modified approach is able to control the false discovery rate and the precision for the whole process.

By repeating the screening process over a large number of randomly chosen subgroups of available subjects, then combining the findings, the proposed method can help to balance out unwanted sources of variation in observational data. Cross-validation through resampling also addresses statistical overfitting \cite{Baily} of the underlying empirical model and enhances the reproducibility of the screening results. The frequency distribution of detection recurrence provides insights into group behavior among significant factors. This methodology  can be used to analyze omics data to identify potential regulators and to explore the inter-relationship between them.  In this paper, we demonstrate the usefulness of our methodology for gene expression data analysis.

The next section provides further background on the empirical Bayes approach for the multiple testing problems. The proposed methodology along with the associated power and precision probability calculations are presented in Section \ref{s_method}. The methodology is illustrated in Section \ref{s_example} using a microarray data set and an RNA-sequencing data set. A few simulation studies are also included in that section along with related power analysis. Some discussion and concluding remarks are given in Section \ref{s_discussion}.  Appendix \ref{app:F_association} presents an argument supporting the association plots developed by the method and Appendix \ref{app:Pooling data}, on pooling data, gives a justification of the methodology and suggests when clusters of hypotheses may act in sync. A performance comparison between the proposed method and an existing screening method is included in Appendix~\ref{app:Performance}.

\section{Background}
\label{s_background}
For a multiple-testing problem,  p-values associated with the hypotheses tests either correspond to the baseline/background information (cases when the null hypothesis is true) or to  the signal (true discoveries, i.e., the cases when the null hypothesis is false and p-values close to 0). The basic model for the density of the population of p-values under study is assumed to be \eqref{eq:main_model} where we label the background distribution $f_0$ as \textbf{\textit{null }} and signal distribution $f_1$ as \textbf{\textit{non-null}}. For any continuous test statistic, the p-value will follow a Uniform distribution under the null hypothesis \cite{Dickhaus}, whereas the Beta distribution can be a natural choice for the non-null (signal) p-values because it is flexible enough to capture distributions with a wide range of shapes on $[0,1]$.

For an observed p-value $x$, the posterior probabilities $P(null|x)$ and $P(non\mbox{-}null|x)$ can be given by the assignment functions associated with the mixture model in \eqref{eq:main_model}: $A_0=p_0f_0(x)/f(x)$  and $A_1=p_1f_1(x)/f(x)$. In an empirical Bayes approach, $A_0$ is estimated using $p_0$, $f_0$ and $f$  fitted from the observed data and for the multiple testing scenario, it is referred to as the \textbf{\textit{local false discovery rate (local fdr) }} (see \cite{Efron07,Efron08,Efron10}).

The local fdr can be used to calculate the \textbf{\textit{tail-area false discovery rate}} $Fdr(x) = P(\bigl. null\Bigl|\Bigr.\  |X| > |x|\bigr)$ (for symmetric $f$ in two-sided tests; can be adjusted for non-symmetric cases). The tail-area Fdr is a useful tool to screen for potentially significant cases. Specifically, observations with small Fdr can be viewed as less likely to be a false discovery and thus can be considered as a significant case or a true discovery. (See \cite{Efron07} for a detailed discussion of local and tail-area false discovery rates and their relationship with the Benjamini-Hochberg \cite{Benjamini}  (BH) mode of controlling overall false discovery rate).

In mixture model based Fdr screening \cite{Efron07, Efron08, Murlidharan}, generally, the two-class model is fitted first using the whole data and then it is used for Fdr based screening to detect the signal or the significant cases. The procedure is equivalent to a classification problem where the estimated Fdr is used as the discriminant function. Estimation of the Fdr is based on the fitted model that is entirely data-driven, and hence prone to overfitting.

Statistical overfitting \cite{Baily} is an especially important issue in machine learning and observational biological data analysis where empirical model building remains essential. Usually, a prognostic model is developed based on the available data, and then the model is used to classify factors under study or to predict the factors' behavior in general. The problem arises when the developed model might work on the given dataset but fails to perform correctly in another data set thus creating a potential reproducibility issue \cite{Baily, Raeder}.

Cross-validation is recommended as a common remedy to avoid overfitting due to data-driven empirical model building \cite{BERRAR, Harell, Mathur, Raeder, Simon, Subramanian}. However, often additional data to validate the fitted model are not available, and the entire analysis or inference is to be done using only one observed data set. Resampling methods are the natural choice for cross-validation in such scenarios.

The proposed method uses the mixture model-based false discovery rate control described above with the following key attributes:
\begin{itemize}
\item We incorporate a random sample splitting step in the existing multiple hypothesis testing procedure with Fdr screening where one part of the data is used for model fitting, and the complementary part is used to screen for the potential signal. This fitting/screening step is used repeatedly with multiple resampled subsets of all available subjects. The most frequently detected factors over all subsets are considered statistically significant. The cross-validation through repeated subsampling and screening helps to reduce the effect of possible overfitting.
\item Observational studies that are common in biology are prone to include many latent sources of variation that can attenuate or distort the signal of interest. The proposed method screens a large number of different data splits and then selects significant cases taking into account the variations over all these splits. This helps to balance out the unwanted sources of variation.
\item We offer a novel way to control the overall false discovery of the inference process by choosing the critical detection frequency of a factor to be identified as a potential true discovery.
\item Based on the resampled screening frequencies, we also present an association plot  that can be used to explore the inter-relationship between the significant factors.
\end{itemize}

\section{Methodology}
\label{s_method}
We now consider a multiple testing scenario where null hypotheses $H_{0,1}, H_{0,2}\ldots, H_{0,M}$  are to be tested against appropriate alternatives respectively. Suppose $T_1, T_2, \dots, T_M$ are test statistics to be used for individual tests considering the available sample size. To identify the non-null cases we can analyze either the p-values associated with each $T_i$ or the left-tail area (LTA) from each $T_i$. Non-null cases will produce p-values close to 0, equivalently non-null LTA's will be close to 0 or 1. The advantage of working with LTA's is one can easily identify the direction of deviation from the null hypothesis.

Let $X_i$ denote the p-value (or LTA) from test statistic $T_i$. When the $X_i$'s follow a two-point mixture distribution with density \eqref{eq:main_model} the proposed methodology is implemented by the following steps.

\textbf{\underline{Proposed Sample-Splitting Analysis Methodology Steps:}}
\medskip
 \begin{description}
 \item [ (i)] First the subjects under study are randomly split into two (equal) parts, \textit{viz}. the \textbf{``\textit{modeling}"} set and the \textbf{``\textit{screening}"} set. The p-values or the LTA's from the test statistic associated with each of the hypotheses under study derived from the modeling set are, respectively called the \textit{\textbf{``modeling data"}} and  the \textit{\textbf{``screening data"}}.
 \item[(ii)] Then, a mixture contamination model $\hat{f}(x)=\hat{p}_0\hat{f}_0(x)+\hat{p}_1\hat{f}_1(x)$ is fitted  \textbf{using only the modeling data}. This fitted model is then adjusted to an ``empirical model" that better captures the baseline and the signal in the empirical fit more appropriately.
  \end{description}
  The detection of significant cases is based on the\textbf{ screening data alone}.
  \begin{description}
   \item[ (iii)] The fitted model is used to derive the tail-area Fdr or the local fdr associated with each $x_i$ in the screening split. The cases with the tail-area Fdr (or the local fdr) less than a predetermined cutoff point are identified as potential significant cases.
   \item[ (iv)] The entire process (stages i, ii, iii) is repeated several times with different random splits of the modeling and the screening subsets. For each split/repetition, a set of potential significant cases is identified. The most frequently identified significant cases are considered as ``significant discoveries". The required critical detection frequency can be set so that the overall false discovery rate for the  process is controlled at a fixed level.
   \item[(v)] The screened cases detected together and their detection frequencies can be used to study the inter-relationships/dependencies between the significant cases.  This frequency distribution is used to develop an association plot for the hypotheses that graphically describes these insights.\\
\end{description}

Next we present the theoretical model formulation along with the screening and power analysis tools. As noted earlier in Section \ref{s_intro}, the local false discovery rate (fdr) in \cite{Efron07} is essentially the same as the assignment function $A_0$ . From identity (2.15) in \cite{Efron10}, the relationship between the local fdr and the tail-area $Fdr(B)$ for a given \emph{tail-area} $B$ is  $Fdr(B) = E(fdr(X)|X\in B)$.

\subsection{Empirical Fit}
\label{subs_emp_fit}

 Using the observed $X_i$'s in the modeling split, we first fit  a mixture of \textit{Uniform} distribution $f_0^*$ and \textit{Beta} distribution $f_1^*$  to the population density given by \eqref{eq:main_model} {\it viz.}$$\hat{f}(x)=p_0^*\cdot f_0^*(x)+p_1^*\cdot f_1^*(x).$$
Since our main interest is the identification of the most extreme cases, we adjust the signal/contamination part as follows: let $f_1^*=f^*_{01}+f^*_{11}$, where

\begin{subequations}
\label{eq:f_cutoff}
\begin{equation}
f^*_{01}(x)=f_1^*(x)\cdot I\{f^*_1(x)<1\}+1\cdot I\{f_1^*(x)>1\}
\end{equation}
\begin{equation}
f^*_{11}(x)=0\cdot I\{f^*_1(x)<1\}+[f^*_1(x)-1]\cdot I\{f_1^*(x)>1\}.
\end{equation}
\end{subequations}
So, $f^*_{11}$ captures the more extreme part of the signal (deviation from the $Uniform$). Since $f^*_{11}$ is not a density, it needs to be normalized as follows. Let $\int_\mathbb{R} f^*_{11}(x)dx=A_{11}$ and define
\[\hat{f}_1(x)=\frac{1}{A_{11}} f^*_{11}(x)\]
and
\[\hat{p}_1=p_1^*\cdot A_{11}\ \mbox{ with }\  \hat{p}_0=1-\hat{p}_1.\]
Now let
\[\hat{f}_0(x)=\frac{p^*_0}{p_0}\cdot f_0^*(x)+\frac{p_1^*}{p_0}\cdot f^*_{01}(x).\]
Then the fitted model can be re-written as
\begin{equation}
\label{eq:fit}
\hat{f}(x)=\hat{p}_0\hat{f}_0(x)+\hat{p}_1\hat{f}_1(x).
\end{equation}
Note that the fitted mixture model is unchanged; the terms have been rearranged so that $\hat{f}_0$ captures more of the middle part of the data while $\hat{f}_1$ captures the tail part. The rearrangement given in \eqref{eq:fit} is what we will refer to as the ``empirical mixture model" and is related to Efron's ``empirical null" \cite{Efron07}; this representation of the fitted mixture model better captures the baseline and the signal distribution than the original $Uniform/Beta$ mixture representation. To derive the estimates for expressions presented in next two subsections one has only to replace the subsequent terms in the assumed population density $f(x)=p_0f_0(x)+p_1f_1(x)$ by Equation \eqref{eq:fit}.

 \textbf{Comment:}
The assumption that the null data follows a \textit{Uniform} distribution  may not always hold in discrete cases (see \cite{muralidharan2012detecting}). But the tail adjustment part  can compensate, at least in part, for the deviation from Uniform in the final adjusted form $\hat{f}_0$.

\subsection{Screening Significant Cases Based on Fdr and Sample Splitting.}
\label{subs_screening}
In case p-values are used for the analysis, $X_i$'s close to $0$ are associated with the signal. If $X_i$'s are the LTA's from the test statistics, then $X_i$'s close to $0$ or $1$ (or both) are associated with the signal depending on whether a left-sided test, right sided-test, or two-sided test is being used. Thus the tail area $B$ (left, right or two-sided) used for calculating tail-area Fdr will depend on the definition of the $X_i$'s and the direction of the alternative hypotheses under study.

Let $F$ be the distribution function corresponding to $f$.  Define $F(B)=\int_Bf(x)dx$, for any Borel set $B$. Similarly, write $F_0$ as the distribution function of the baseline distribution $f_0$.  With this notation, we derive the tail-area Fdr associated with a given tail-area $B$ as follows:
\begin{align}
\label{Efdr}
Fdr(B)&=E(fdr(X)\mid X\in B)=\int_B\frac{fdr(x)dF(x)}{F(B)}\nonumber\\
&=\frac{1}{F(B)}\int_B\frac{p_0f_0 (x)}{f(x)}\cdot f(x)dx=\frac{1}{F(B)}\int_Bp_0f_0(x)dx=\frac{p_0F_0 (B)}{F(B)}.
\end{align}
In practice, one can consider the tail-area $B(x)$ associated with any value $x$  in the support $\mathbb{S}$ of $f$. The tail-area false discovery rate $Fdr\left(B(x)\right)$ associated with $x$ can be calculated using \eqref{Efdr}.

 If $X_i$'s are p-values derived for each hypothesis under study, for any $x\in\mathbb{S}$, the appropriate tail area to use is  $B(x)=\{y\in\mathbb{S}:y< x\}$. When $X_i$'s are LTA's from test statistics, one should use $B(x)=\{y\in\mathbb{S}:y< x\}$ for a left sided test, and $B(x)=\{y\in\mathbb{S}:y> x\}$ for a right sided test. Whereas, in a two sided test situation with $f$  symmetric around zero,  the tail area simply is $B(x)=\{y\in\mathbb{S}:|y|>|x|\}$.

 In general, for $f$  that is not symmetric, the two-sided tail area can be derived using matching percentiles. Express any $x$  as the $p^{th}$ percentile of $f$, i.e, $\int_{-\infty}^{x}f(u)du=p$; then a matching $x^*$ can be found such that $\int_{-\infty}^{x^*}f(u)du=1-p$. Now if $p<0.5$ ($x$ is smaller than the median) we choose
\begin{subequations}
 \label{eq:tailarea}
 \begin{equation}
 B(x)=\{y\in\mathbb{S}:y<x\}\cup\{y\in\mathbb{S}:y>x^*\}.
 \end{equation}
  On the other hand if $p>0.5$ ($x$ is larger than the median) we choose
  \begin{equation}
  B(x)=\{y\in\mathbb{S}: y<x^*\}\cup\{y\in\mathbb{S}:y>x\}.
  \end{equation}
\end{subequations}
An observation $x$ with $Fdr\left(B(x)\right)$ smaller than a predetermined critical value should be identified as significant.

For tail-area Fdr screening of a data set,  a modeling split is first used to fit the adjusted mixture model \eqref{eq:fit}. Then for each data point $x_i$ in the corresponding screening split, the appropriate tail area $\hat{B}(x_i)$ is determined using the fitted model. Next $\hat{f}_0$, $\hat{p}_0$ and $\hat{f}$ from the modeling fit are used with \eqref{Efdr} to derive the estimated observed tail-area Fdr, \emph{viz}. $\widehat{Fdr}(\hat{B}(x_i))$. Any case with $\widehat{Fdr}(\hat{B}(x_i))<q$ is screened as a potential discovery, where $q$ is a pre-determined cutoff point.

For screening with local fdr after fitting the adjusted mixture model with the modeling split, the fitted densities are used to calculate estimated local fdr $\widehat{fdr}(x_i)=\frac{\hat{p}_0\hat{f}_0(x_i)}{\hat{f}(x_i)}$ for each screening split data point. Cases with $\widehat{fdr}(x_i)$ less than $q$ (predetermined) are considered to be potential discoveries.

 In terms of the rejection region, for any fixed cutoff point $q$, depending on the tail-area Fdr or the local fdr screening,  the theoretical rejection set from the $k^{th}$ split is given by
 \begin{subequations}
 \label{eq:R}
\begin{equation}
\label{Rq}
R_k(q):=\{x\in\mathbb{S}_k: \widehat{Fdr}\left(B(x)\right)<q\}
\end{equation}
or
\begin{equation}
\label{tRq}
\tilde{R}_k(q):=\{x\in\mathbb{S}_k: \widehat{fdr}\left(B(x)\right)<q\},
\end{equation}
\end{subequations}
where $\mathbb{S}_k$ is the support of $\hat{f}$ from the $k^{th}$ sample split.

The above calculation is repeated a number of times. The potential significant cases can be identified from the combined rejection set $\bigcup_kR_k(q)$ or $\bigcup_k\tilde{R}_k(q)$. But to increase the precision, only the observations that have been detected repeatedly with high frequency across the $R_k(q)$'s or $\tilde{R}_k(q)$'s should be considered as potential true discoveries. The critical frequency of detection for significant discoveries will be discussed in sub-Section \ref{subs:power} (see Equation \eqref{eq:freq_cutoff_N}).

\subsection{Power and Error Probabilities Calculation.}
\label{subs:power}
 Efron \cite{Efron07, Efron10} uses a whole data fit on z-transformations of p-values associated with the hypotheses under study and advocates the use of the local fdr for the screening of significant cases. In that analysis, for a given cutoff point $q$ of the local fdr, the rejection region effectively is $\tilde{R}(q)=\{x: fdr(x)<q\}$. The power diagnostic tools chosen in those discussions are the non-null average of the local fdr  $E_{H_1}(fdr)$  and the non-null cdf of the local fdr given by $G(q)=P_{H_1}(fdr<q)=P_{H_1}\left(\tilde{R}(q)\right)=\int_{\tilde{R}(q)}f_1(x)dx$. Some empirical estimates of these functions were used in \cite{Efron07, Efron10} for the power analysis.

For the sample splitting method proposed in this paper, where the model is fitted on the p-values or LTA's associated with test statistics, the rejection region from the $k^{th}$ split can be obtained from \eqref{eq:R}. The combined rejection region from all splits can then be constructed as $R(q)=\bigcup_kR_k(q)$ or as $R(q)=\bigcup_k\tilde{R}_k(q)$, depending on the screening tool used. Considering the mixture model setup in \eqref{eq:main_model}, for a given rejection region $R(q)$ with a cutoff point $q$, the following probabilities can be used for the power analysis and a relative efficiency comparison:

\begin{equation}
\label{recall}
\mbox{\textbf{\textit{Power}}}=G^*(q)=P_{H_1}\left(R(q)\right)=\int\limits_{R(q)}f_1(x)dx \mbox{ (also known as \textbf{\textit{recall }}or \textbf{\textit{sensitivity}})}.
\end{equation}
\begin{equation}
\label{typeI}
\mbox{\textbf{\textit{Type I error}}}=P_{H_0}\left(R(q)\right)=\int\limits_{R(q)}f_0(x)dx \mbox{ (also known as \textbf{\textit{false positive rate}})}.
\end{equation}
\begin{equation}
\label{typeII}
\mbox{\textbf{\textit{Type II error}}}=P_{H_1}\left(R^c(q)\right)=\int\limits_{R^c(q)}f_1(x)dx \mbox{ (also known as \textbf{\textit{false negative rate}})}.
\end{equation}
\begin{equation}
\label{precision}
\mbox{\textbf{\textit{Precision}}}=P\left(H_1|R(q)\right)=\dfrac{p_1\int\_{R(q)}f_1(x)dx}{\int_{R(q)}f(x)dx}.
\end{equation}
\begin{equation}
\label{Fdr_o}
\mbox{\textbf{\textit{Overall  combined false discovery rate, }}} \mathbf{Fdr_O}=P\left(H_0|R(q)\right)=\dfrac{p_0\int\limits_{R(q)}f_0(x)dx}{\int\limits_{R(q)}f(x)dx}.
\end{equation}
Here, $f_0$, $f_1$ and $f$  are the true densities that follow from the assumption that $X_1,\ldots,X_n$ are $i.i.d.$ with pdf \eqref{eq:main_model}. The terms ``recall", ``false negative rate" or ``false positive rate" and ``precision" are commonly used in machine learning (see \cite{Powers}). $Fdr_O=1-Precision$ measures the process-wise false discovery rate associated with the combined rejection region $R(q)$.

The estimates of \cref{recall,typeI,typeII,precision,Fdr_o} for a given data set can be obtained from the following steps:\medskip\\
\textbf{\underline{Rejection Region and Power Calculation Steps.}}
\begin{description}
\item[(i)] Suppose sample splitting and subsequent screening were done $N$ times following the steps in Section \ref{subs_screening} and for a given $q$ the rejection regions $R_k(q)$  or $\tilde{R}_k(q)$ (as in \eqref{eq:R}) were obtained from each of the splits.
\item[(ii)] For a two-sided test with LTA's the rejection region from the $k^{th}$ split, using either the tail-area Fdr or the local fdr screening, can be written as:
\begin{align*}R_k(q)&=\{x:Fdr\left(B(x)\right)<q\}=\{x<x_k^*\}\cup\{x>x_k^{**}\}\\
\tilde{R}_k(q)&=\{x: fdr\left(B(x)\right)<q\}=\{x<\tilde{x}_k^*\}\cup\{x>\tilde{x}_k^{**}\}.
\end{align*}
Then writing
\begin{align*}
x^*=\max\limits_{k}x_k^*\mbox{ and }&\; x^{**}=\min\limits_{k}x_k^{**},\\
 \tilde{x}^*=\max\limits_{k}\tilde{x}_k^*\mbox{ and }&\;\tilde{x}^{**}=\min\limits_{k}\tilde{x}_k^{**},
\end{align*}
the combined rejection region can be expressed as:
\begin{subequations}
\label{R_comb}
\begin{equation}
\label{R_comb1}
R(q)=\bigcup_{k=1}^NR_k(q)=\{x<x^*\}\cup\{x>x^{**}\}
\end{equation}
or
\begin{equation}
\label{R_comb2}
R(q)=\bigcup_{k=1}^N\tilde{R}_k(q)=\{x<\tilde{x}^*\}\cup\{x>\tilde{x}^{**}\}
\end{equation}
\end{subequations}
depending on the choice of the screening tool. Equation \eqref{R_comb} will include sets with one-sided regions only for p-value analysis or one-sided tests with LTA's.
\item[(iii)] A mixture model $\tilde{f}(x)=\tilde{p}_0\tilde{f}_0(x)+\tilde{p}_1\tilde{f}_1(x)$ with tail adjustment fitted to the entire data (without any data splitting) can be used for the estimates of the densities in \cref{recall,typeI,typeII,precision,Fdr_o}.
\item[(iv)] Numerical integration and numerical root finding techniques can be used to estimate the probabilities in \cref{recall,typeI,typeII,precision,Fdr_o} and to find $x_k^*$ and $x_k^{**}$ or $\tilde{x}_k^*$ and $\tilde{x}_k^{**}$ from \eqref{eq:R}, where closed form solutions are not feasible.
\end{description}
The power and error probabilities in the steps above are associated with the final analysis method that combines all $N$ splits and not with any single modeling split in particular. Therefore, for the estimation of \cref{recall,typeI,typeII,precision,Fdr_o} it is appropriate to use a mixture model fitted to the entire data for estimates of densities $f_0$, $f_1$ and $f$ as suggested in step (iii) above. An individual fit from any single particular modeling split should not be used for the power analysis.

Using the combined rejection region $R(q)$ for screening will increase the number of rejections compared to whole data-based screening described in \cite{Efron07}. This will naturally increase the power \eqref{recall} of the proposed method, but the payoff will be a loss of precision \eqref{precision} or increase in $Fdr_O$.

\textbf{\underline{Critical Detection Frequency.}} \medskip\\
Using only the cases in $R(q)$ with high detection frequencies as the potential discoveries will increase the precision of the method. If we want to impose a tolerance level  $q^*$ for $Fdr_O$, a critical detection frequency requirement can be derived  as follows:
\begin{align}\label{eq:freq_cutoff}
Fdr_O<q^* &\Rightarrow \dfrac{p_0\int_{R(q)}f_0(x)dx}{\int_{R(q)}f(x)dx}<q^*\nonumber\\
&\Rightarrow \int_{R(q)}f(x)dx>\dfrac{p_0\int_{R(q)}f_0(x)dx}{q^*}\nonumber\\
&\Rightarrow P\bigl[X_i\in R(q)\bigr]>\dfrac{p_0\int_{R(q)}f_0(x)dx}{q^*}.
\end{align}
Note that the $i^{th}$ factor is detected as potentially significant if $x_i\in R(q)$. For any factor (individual hypothesis) under study, the relative frequency of detection from $N$ repeated sample splittings is essentially the bootstrap resampling relative frequency of lying in the rejection region $R(q)$. Thus, for large $N$ the relative frequency of detection for the $i^{th}$ factor estimates $P\bigl[X_i\in R(q)\bigr]$, $i=1,2,\ldots M$.

 Following Equation \eqref{eq:freq_cutoff}, if only factors with relative  frequency of detection larger than $\tilde{p}_0\int_{R(q)}\tilde{f}_0(x)dx/q^*$ are considered to be significant, that will control the estimated $Fdr_O$ at $q^*$. In other words, a factor can be considered a significant discovery  if after $N$ run of sample splitting and subsequent screening \textbf{the frequency of detection is at least}
\begin{equation}
\label{eq:freq_cutoff_N}
N\times\dfrac{\tilde{p}_0\int_{R(q)}\tilde{f}_0(x)dx}{q^*}.
\end{equation}
Here the choice of $q$ and $q^*$  will be user defined (just as the choice of the level of significance for a traditional single hypothesis test). Smaller $q$ and $q^*$ will result in more conservative screening.

An added benefit of expressing the error probabilities as a function of Fdr cutoff point $q$ is that one can choose $q$ where both the type I and  the type II error probabilities are at a reasonable level. Alternatively, an appropriate $q$ can also be chosen so that the proportion of correct classifications $\tilde{A}(q)=p_1F_1\left(R(q)\right)+p_0F_0\left(R^c(q)\right)$ is at a desired level.  $\tilde{A}$ is also known as the ``\textbf{\textit{accuracy}}" function in machine learning.

\section{Illustrative Examples.}
\label{s_example}
In this section, we illustrate the proposed methodology with a microarray data set and an RNA-sequencing data set where the goal is to identify genes that are expressed at a significantly higher or lower level in the experimental group compared to the control group. We also present several simulation studies to compare the proposed approach with the existing Fdr screening methods.

\subsection{Microarray Data Analysis.}
\label{subs:microarray}
Empirical Bayes methods have been used to analyze microarray data \cite{smyth2004linear, hirakawa, Efron08}. The proposed methodology is more generalized since it can be applied with any test statistic (not necessarily Normal or t-distribution based).

We analyzed a prostate cancer microarray data set (used in \cite{Efron10} from \cite{Singh}, the data is available in $\texttt{R}$ under package $\texttt{sda}$, in a data file named ``singh2002") consists of 52 prostate cancer patients and 50 normal subjects. In the data, the expression levels of the same 6033 genes were measured for each subject. The analysis aims to detect genes that have significantly different expression levels between cancer and non-cancer groups and to explore the inter-relation between those genes. The significant genes are captured in the screening step. An association plot is generated based on concurrent detection frequencies,  that is useful for exploring the inter-relation between these genes (see section \ref{s_discussion} for discussion and interpretation of these plots). We used LTA's with tail-area Fdr screening for the analysis, although a local fdr screening also can be used following the steps described in Section \ref{s_method}.

\begin{figure}[!h]
\vspace{-0.4 cm}
\centering
\begin{subfigure}{.5\textwidth}
  \centering
  \includegraphics[width=.9\linewidth]{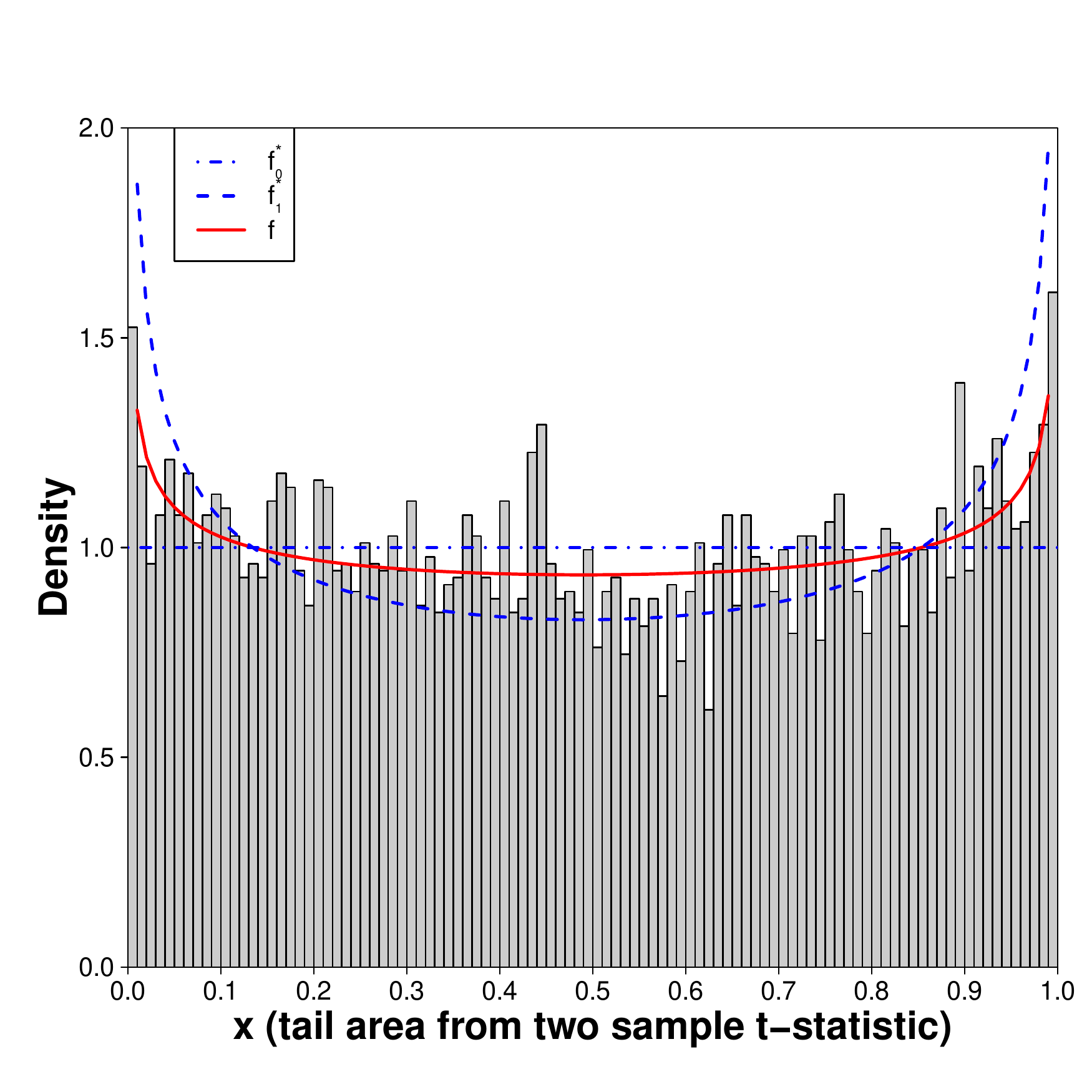}
  \caption{Uniform-Beta mixture distribution.}
  \label{fig: mc_mix1}
\end{subfigure}%
\begin{subfigure}{.5\textwidth}
  \centering
  \includegraphics[width=.9\linewidth]{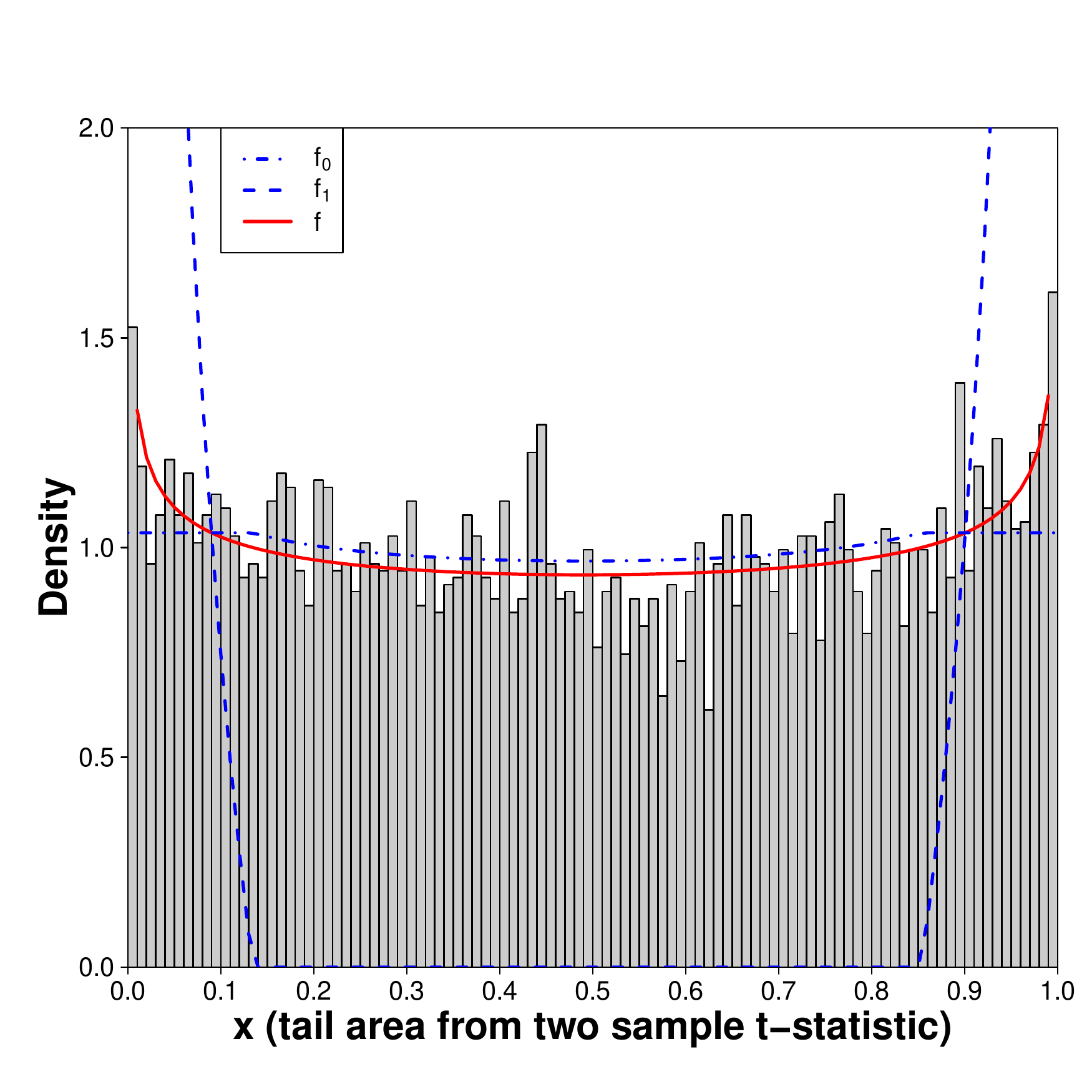}
  \caption{Adjusted Uniform-Beta mixture distribution.}
  \label{fig: mc_mix2}
\end{subfigure}
\medskip
\caption{The histogram of $x_i$ (left-tail-area for the observed two sample t-statistic for genes $i=1,2,\ldots,6033$) from a particular screening split consisting of half of the control and the treatment group respectively. Superimposed on \ref{fig: mc_mix1} are the fitted Uniform-Beta (blue dashed lines) and the associated mixture distribution (red solid line) obtained from the corresponding modeling split as $\hat{f}(x)=0.622f_{0}^{\ast}(x)+0.378f_{1}^{\ast}(x)$ where $f^*_{0}$ is the $\text{Uniform}(0,1)$ pdf and $f^*_{1}$ is the $\text{Beta}(0.696,0.736)$ pdf. Figure~\ref{fig: mc_mix2} is the  empirical null fit adjusted from the fitted Uniform-Beta mixture distribution as in Equation \eqref{eq:fit} where $\hat{f}(x)=0.966\hat{f}_{0}(x)+0.034\hat{f}_{1}(x)$.}
\label{fig: mc_mix}
%\vspace{-0.3 cm}
\end{figure}

To begin, the data was split into the modeling set and the screening set. The modeling split consisted of 26 randomly selected prostate cancer patients and 25 normal subjects.  The remaining 26 patients and 25 non-cancer subjects formed the screening split.  The modeling data was used to fit the contamination model \eqref{eq:fit}.  This is described next.

The two-sample $t$-statistic $t_i$, $i=1, 2,\ldots 6033$, is calculated for each gene from the modeling group where it is assumed that $t_i$ follows a central $t$-distribution with 49 degrees of freedom.  The LTA for each of these $t_i$'s is calculated as:
\begin{equation}
\label{eq:xi}
x_i= P(t<t_i),\; i=1, 2,\ldots 6033.
\end{equation}
Note that, $x_i$'s should be close to 0 or 1 for genes that are deemed significantly differentially expressed between cancer and non-cancer groups. To compute the $t$-statistics, we used the difference (control mean-cancer mean). Hence, any $x_i$ close to $0$ implies that the $i^{th}$ gene is expressed at higher level in cancer patients, while any $x_i$ close to $1$ indicates that the $i^{th}$ gene is expressed at lower level in cancer patients  compared to the control group. The histogram of the $x_i$'s from 6033 genes in  Figure~\ref{fig: mc_mix} shows a bathtub shape. Prompted by that shape, we first fitted a mixture of $Uniform(0,1)$ and a $Beta$ distribution to the modeling data and then readjusted it as described in subsection \ref{subs_emp_fit}.

Figure~\ref{fig: mc_mix}  illustrates the histogram of $x_i$ (left-tail-area for the observed two sample t-statistic for genes $i=1,2,\ldots,6033$) from a particular screening split consisting of half of the control and the treatment group respectively. Panel~\ref{fig: mc_mix1} features the fitted Uniform-Beta (blue dashed lines) and the associated mixture distribution (red solid line) obtained from the corresponding modeling split as $\hat{f}(x)=0.622f_{0}^{\ast}(x)+0.378f_{1}^{\ast}(x)$, where $f^*_{0}$ is the $\text{Uniform}(0,1)$ pdf and $f^*_{1}$ is the $\text{Beta}(0.696,0.736)$ pdf. Panel~\ref{fig: mc_mix2} shows the adjusted empirical null fit as in Equation \eqref{eq:fit} derived as $\hat{f}(x)=0.966\hat{f}_{0}(x)+0.034\hat{f}_{1}(x)$.

Next following \eqref{Efdr}, we computed the tail-area Fdr associated with each gene. Genes with tail-area Fdr less than 0.1 were declared as potentially significantly different between cancer patients and non-cancer subjects. This procedure was repeated on 500 different sample splits.

Out of these 500 repetitions, 234 verification groups identified at least one significant gene with associated tail-area Fdr less than 0.1. The other 266 verification groups failed to capture any significant gene. Few verification sets identified more than one significant gene. In the 500 repetitions of this procedure, out of 6033 total genes, 117 genes showed Fdr$<$ 0.1 at least once, and 47 genes had FDR$<0.1$ at least twice. These genes included some that are expressed at significantly higher levels among the cancer patients ($x_i$'s close to 0) and some at significantly lower levels among the cancer patients ($x_i$'s close to 1).

Genes that are repeatedly detected as significant  strongly confirm the difference between the patient and the control group. Table~\ref{table_gene_freq1} shows the significant genes (with detection frequency at least 2) along with the number of times they were identified as significant through the 500 sample splits. The false discovery rate calculated from the whole data analysis by the BH \cite{Benjamini} method is also included in the table ($\mathrm{FDR_{BH}}$).

\begin{center}
\begin{longtable}{crrrrrrr}
\caption{The prostate cancer data was analyzed using the proposed method with 500 random sample splits. Our method discovered 47 genes by using the Fdr threshold 0.1, and the detection frequency threshold 0.004 (at least two significance detection occur), respectively. Columns freq and rfreq show the detection frequency and relative frequency respectively $\mathrm{FDR_{BH}}$ represents the Benjamini-Hochberg FDR values by using the whole data for these 47 genes. For each gene, med(Fdr) shows the median tail-area Fdr from 500 screening splits. The columns med($x$), ave($x$) and sd($x$) are the median, mean and standard deviation of the 500 LTA's $x$  for each gene computed from 500 screening splits.}
\label{table_gene_freq1}
\medskip
\\
\hline
Gene\_ID & freq & rfreq & $\mathrm{FDR_{BH}}$ & med(Fdr) &  med(x) & mean(x) & sd(x)\\
\hline
\endfirsthead

\multicolumn{7}{c}
{{\bfseries \tablename \ \thetable{} -- continued from previous page}}\\
\hline
Gene\_ID & freq & rfreq & $\mathrm{FDR_{BH}}$ & med(FDR) &  med(x) & mean(x) & sd(x)\\
\hline
\endhead

\hline
\multicolumn{7}{r}{continued on next page}\\
\hline
\endfoot

\hline
\endlastfoot
610 & 54 & 0.108 & 0.001 & 0.068 & 0.000 & 0.001 & 0.003 \\
  1720 & 51 & 0.102 & 0.005 & 0.063 & 0.000 & 0.002 & 0.006 \\
  4331 & 28 & 0.056 & 0.019 & 0.067 & 0.998 & 0.992 & 0.017 \\
  3940 & 22 & 0.044 & 0.015 & 0.065 & 0.999 & 0.994 & 0.013 \\
  914 & 20 & 0.040 & 0.015 & 0.067 & 0.001 & 0.005 & 0.010 \\
  364 & 15 & 0.030 & 0.015 & 0.076 & 0.999 & 0.995 & 0.010 \\
  4546 & 8 & 0.016 & 0.018 & 0.043 & 0.999 & 0.994 & 0.014 \\
  921 & 8 & 0.016 & 0.060 & 0.083 & 0.994 & 0.985 & 0.027 \\
  3647 & 7 & 0.014 & 0.019 & 0.071 & 0.002 & 0.008 & 0.020 \\
  4316 & 7 & 0.014 & 0.037 & 0.082 & 0.997 & 0.988 & 0.023 \\
  1077 & 6 & 0.012 & 0.031 & 0.066 & 0.002 & 0.011 & 0.023 \\
  2856 & 5 & 0.010 & 0.064 & 0.056 & 0.994 & 0.979 & 0.039 \\
  579 & 5 & 0.010 & 0.019 & 0.081 & 0.002 & 0.007 & 0.012 \\
  1089 & 4 & 0.008 & 0.019 & 0.059 & 0.002 & 0.009 & 0.020 \\
  11 & 4 & 0.008 & 0.092 & 0.041 & 0.011 & 0.030 & 0.048 \\
  1346 & 4 & 0.008 & 0.060 & 0.058 & 0.995 & 0.982 & 0.030 \\
  2 & 4 & 0.008 & 0.063 & 0.079 & 0.006 & 0.018 & 0.032 \\
  2852 & 4 & 0.008 & 0.151 & 0.060 & 0.014 & 0.038 & 0.059 \\
  2945 & 4 & 0.008 & 0.065 & 0.085 & 0.994 & 0.982 & 0.031 \\
  3930 & 4 & 0.008 & 0.070 & 0.078 & 0.008 & 0.023 & 0.041 \\
  4088 & 4 & 0.008 & 0.035 & 0.065 & 0.997 & 0.989 & 0.022 \\
  702 & 4 & 0.008 & 0.063 & 0.090 & 0.995 & 0.983 & 0.029 \\
  3017 & 3 & 0.006 & 0.070 & 0.059 & 0.994 & 0.982 & 0.029 \\
  3991 & 3 & 0.006 & 0.035 & 0.079 & 0.996 & 0.988 & 0.022 \\
  4073 & 3 & 0.006 & 0.040 & 0.065 & 0.997 & 0.987 & 0.028 \\
  4396 & 3 & 0.006 & 0.070 & 0.065 & 0.992 & 0.975 & 0.044 \\
  1068 & 2 & 0.004 & 0.019 & 0.087 & 0.002 & 0.005 & 0.011 \\
  1314 & 2 & 0.004 & 0.060 & 0.059 & 0.005 & 0.015 & 0.026 \\
  1491 & 2 & 0.004 & 0.119 & 0.048 & 0.013 & 0.029 & 0.043 \\
  1588 & 2 & 0.004 & 0.070 & 0.080 & 0.007 & 0.021 & 0.040 \\
  1966 & 2 & 0.004 & 0.120 & 0.071 & 0.014 & 0.036 & 0.055 \\
  2370 & 2 & 0.004 & 0.063 & 0.062 & 0.006 & 0.018 & 0.034 \\
  2621 & 2 & 0.004 & 0.512 & 0.078 & 0.062 & 0.097 & 0.110 \\
  2745 & 2 & 0.004 & 0.627 & 0.080 & 0.912 & 0.862 & 0.145 \\
  2785 & 2 & 0.004 & 0.327 & 0.077 & 0.969 & 0.936 & 0.084 \\
  2912 & 2 & 0.004 & 0.137 & 0.076 & 0.014 & 0.035 & 0.055 \\
  332 & 2 & 0.004 & 0.015 & 0.094 & 0.001 & 0.004 & 0.007 \\
  3600 & 2 & 0.004 & 0.065 & 0.056 & 0.994 & 0.984 & 0.027 \\
  3848 & 2 & 0.004 & 0.386 & 0.064 & 0.040 & 0.074 & 0.092 \\
  4000 & 2 & 0.004 & 0.063 & 0.079 & 0.993 & 0.982 & 0.032 \\
  4040 & 2 & 0.004 & 0.095 & 0.082 & 0.989 & 0.974 & 0.041 \\
  4104 & 2 & 0.004 & 0.060 & 0.086 & 0.994 & 0.982 & 0.031 \\
  4417 & 2 & 0.004 & 0.529 & 0.063 & 0.931 & 0.891 & 0.120 \\
  4515 & 2 & 0.004 & 0.117 & 0.057 & 0.987 & 0.969 & 0.054 \\
  4518 & 2 & 0.004 & 0.033 & 0.085 & 0.003 & 0.010 & 0.021 \\
  4549 & 2 & 0.004 & 0.060 & 0.083 & 0.005 & 0.016 & 0.033 \\
  758 & 2 & 0.004 & 0.255 & 0.067 & 0.973 & 0.946 & 0.071 \\
\hline
\end{longtable}
\vspace{-0.8 cm}
\end{center}

\begin{figure}[!h]
\centering
\vspace{-2 cm}
\begin{subfigure}{.8\textwidth}
  \centering
  \includegraphics[width=0.95\linewidth]{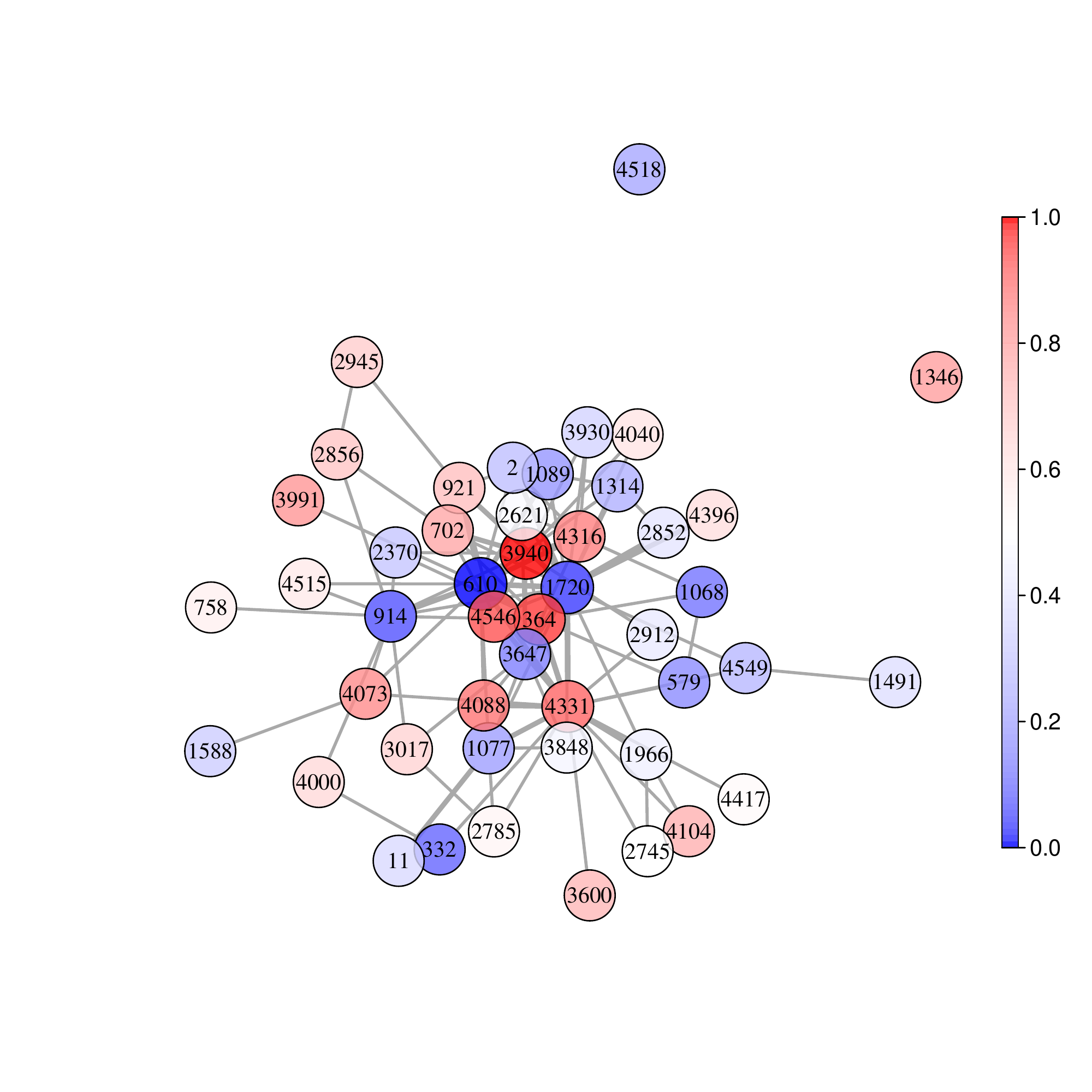}
  \vspace{-1.8 cm}
  \caption{Frequency network for the 47 dietected genes with relative frequency greater than 0.004, by using the tail-area Fdr threshold 0.1.}
  \label{fig: mc_network1}
\end{subfigure}%
\par
\vspace{-1.5 cm}
\begin{subfigure}{.8\textwidth}
  \centering
  \includegraphics[width=\linewidth]{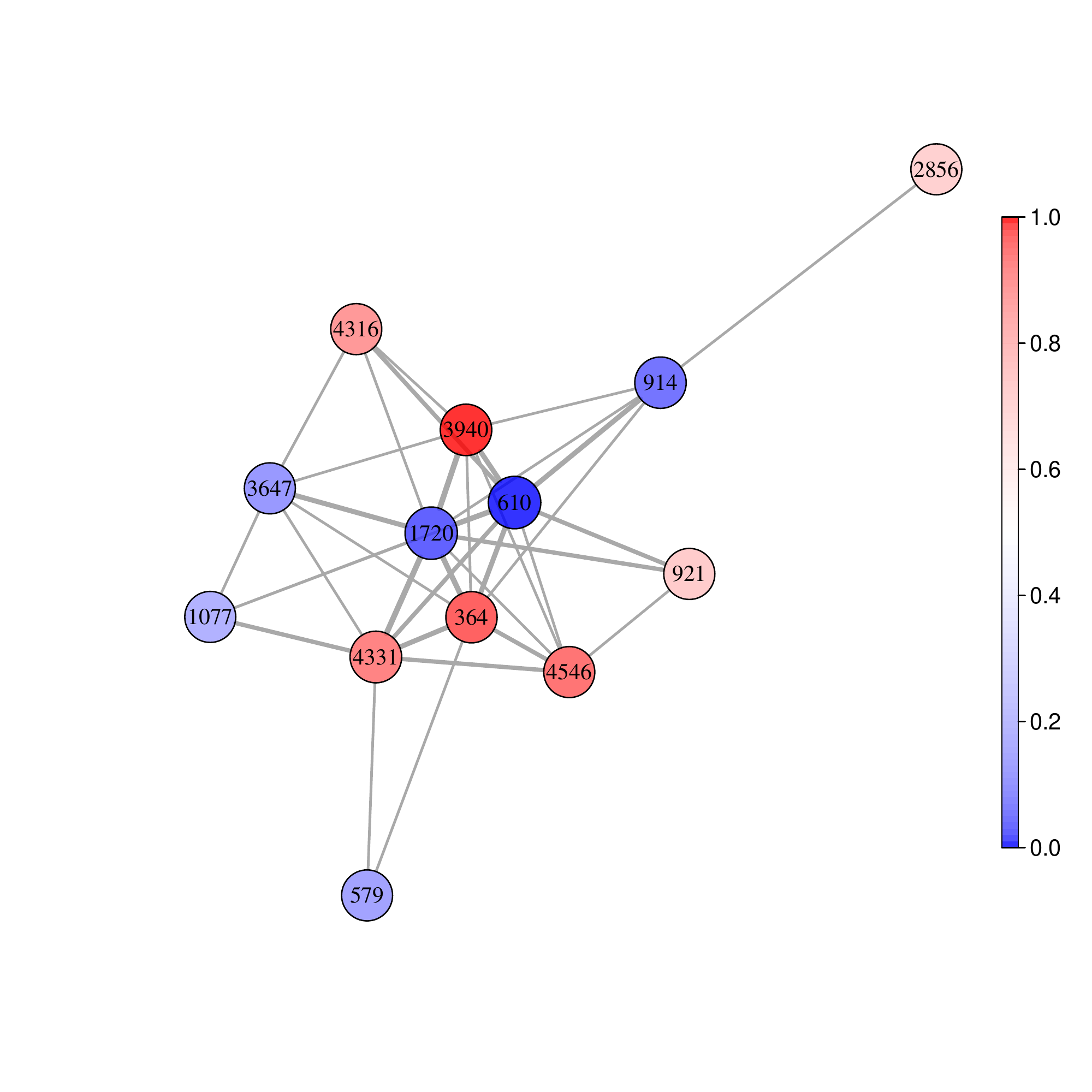}
  \vspace{-2 cm}
  \caption{Frequency network for the  genes with relative detection frequency greater than 0.01.}
  \label{fig: mc_network2}
\end{subfigure}
\medskip
\caption{Frequency network plot from the prostate cancer data analysis. The edge width indicates the frequency of the pair genes detected as significant at the same time. The node color represents the direction of difference in gene expression. That is, the blue nodes represent  genes expressed at significantly higher levels, and the red nodes represent  genes expressed at significantly lower levels in the cancer patients compared to the control group.}
\label{fig: network}
\vspace{-0.5 cm}
\end{figure}

Table~\ref{table_gene_freq1} further shows that the most frequently screened genes by the proposed method would also be detected by the BH method, since they had $\mathrm{FDR_{BH}}<0.1$ (Efron's approach would have a similar result as the BH method since they are related \cite{Efron07}). However, our approach was able to capture a few genes (with low detection frequencies) that would have been missed by the BH screening if the same threshold value 0.1 was used. For example, genes 3848, 4417, 4515, 758 all had $\mathrm{FDR_{BH}}$ larger than 0.1 but was detected twice by our method. That is, these genes showed tail-area $Fdr<0.1$ in 2 screening splits among the 500 splits used in the analysis. Note that, the median LTA ($med(x)$) for these genes are fairly close to 0 or 1 which supports the fact that these are potentially significant genes.

Figure~\ref{fig: mc_network1} shows genes identified as significant at least twice and the sets of genes with which they are simultaneously identified as significant across 500 sample splits and subsequent screenings. In this \textbf{\textit{gene frequency association}} plot (\textbf{\textit{F-association}}), the nodes and edges respectively indicate the detected significant genes and the detection of two genes at the same time in a particular split. The edge width increases with the increase in the frequency of simultaneous detection for a pair of significant genes.  The node color represents the median tail area from the 500 screening data sets for that gene, for which the color turns from blue to red accordingly as the tail area increases from small (close to 0) to large (close to 1). That is, the blue nodes represent  genes expressed at significantly higher levels, and the red nodes represent  genes expressed at significantly lower levels in the cancer patients' group compared to the controls. Figure~\ref{fig: mc_network2} shows genes that are detected at least 5 times for a clearer picture into the network.

\subsection{RNA-seq Data Analysis}

In this subsection, we show the application of our methodology for the analysis of global gene expression data using RNA-seq, a commonly used tool for transcriptome profiling. Here we demonstrate how the method can be used with p-values. We used the data consists of RNA-Seq profiles of cell lines derived from lymphoblastoid cells from 69 different Yoruba individuals from Ibadan, Nigeria \cite{pickrell2010understanding}. The analysis aims to investigate differentially expressed genes between males and females. The RNA counts are available in the \texttt{R} package \texttt{tweeDEseqCountData}. In the raw RNA counts, there are 38415 genes with or without defined annotations. To filter out the noninformative genes, we keep genes with both defined annotations and at least 1 count-per-million (CPM) in at least 20 samples. At last, 17310 genes remain for the differential gene analysis.

\begin{figure}[!h]
\centering
\begin{subfigure}[b]{0.45\textwidth}
\vskip -.2in
\includegraphics[width=\textwidth]{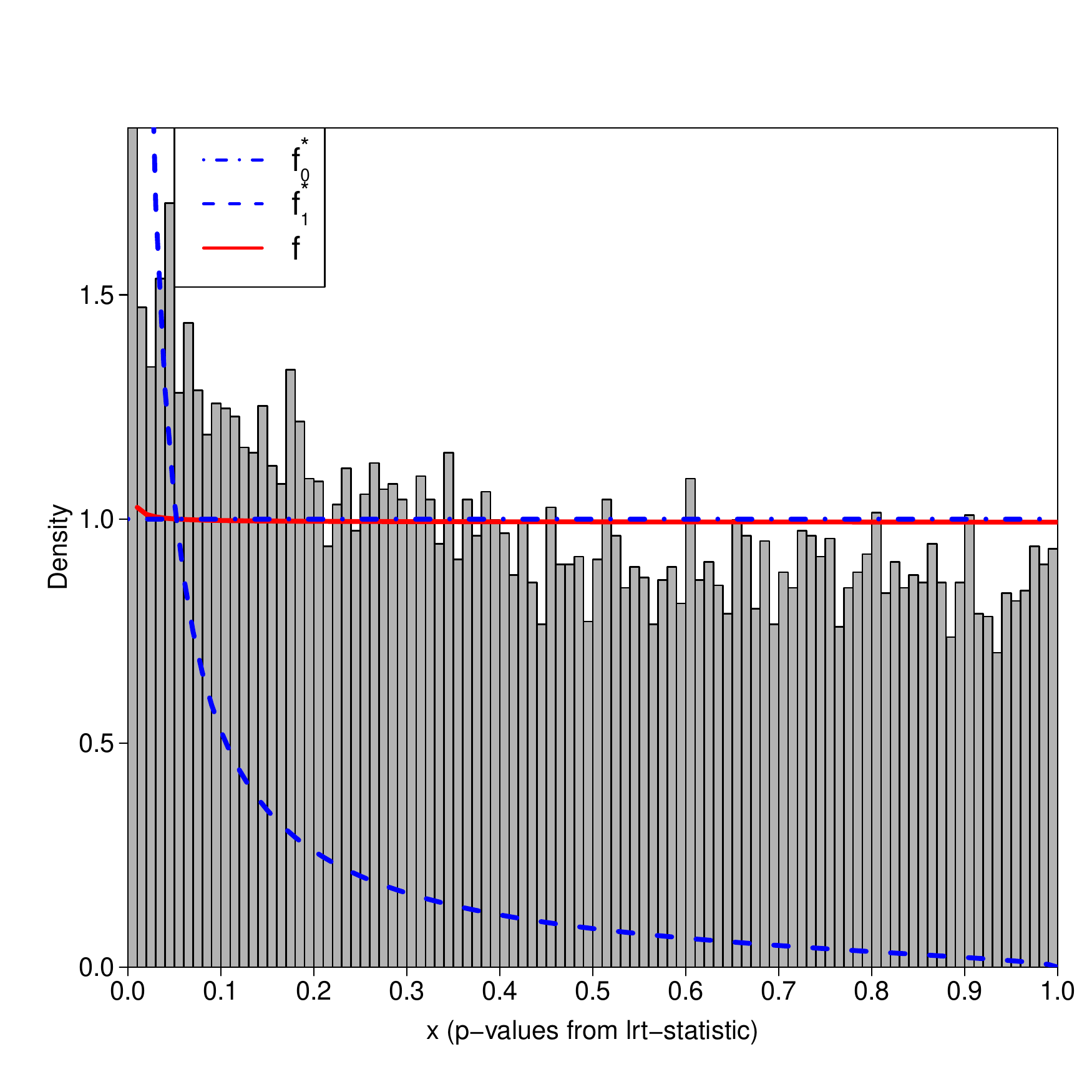}
\caption{Unadjusted fit for the whole data.}
\label{fig: whole1}
\end{subfigure}
%\hfill
\begin{subfigure}[b]{0.45\textwidth}
\vskip -.2in
\includegraphics[width=\textwidth]{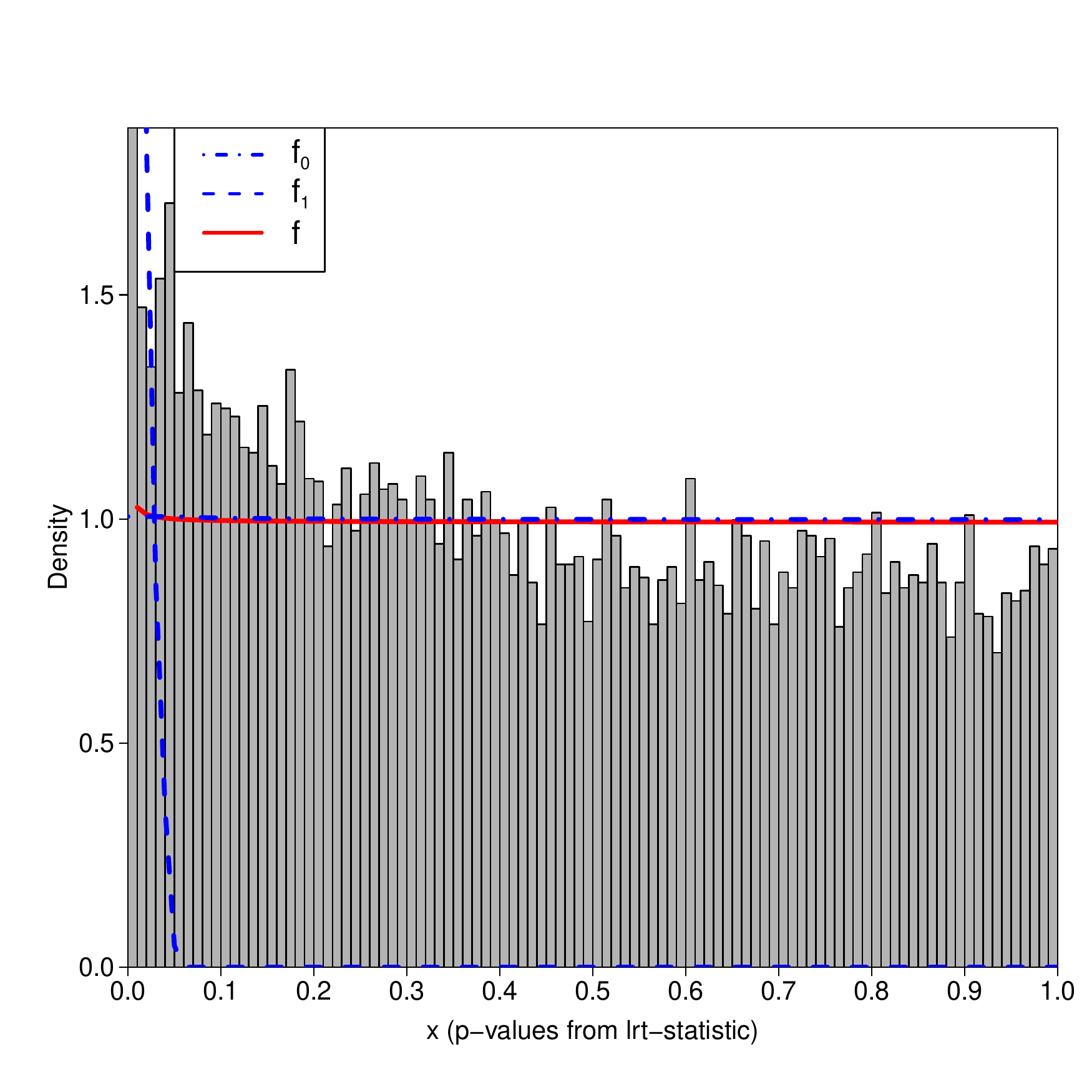}
\caption{Adjusted fit for the whole data.}
\label{fig: whole2}
\end{subfigure}
\medskip
\caption{The mixture model for the p-values using the whole data set. $f(x)=0.993f_0^*(x) + 0.007f_1^*(x)$ where $f_0^*(x)$ is $Uniform(0,1)$ density, and $f_1^*(x)$ is the $Beta(\alpha=0.064,\beta=1.517)$ density. The adjusted mixture model is $f(x)=0.994\hat{f}_0(x) + 0.006\hat{f}_1(x)$.}
\label{fig:whole_pvalue}
\vspace{-0.3 cm}
\end{figure}

\begin{table}[!h]
\begin{minipage}[c]{0.95\linewidth}
\centering
%\hspace{-2.3 cm}
\caption{RNA-seq data analysis with 100 sample splits. Table shows 28 most frequently detected significant genes with tail-area Fdr$<0.01$ (detection frequencies 40 or higher). The ``freq" column indicates the frequency of occurrence for the corresponding gene in the 100 screening splits. The FDR$_{\mbox{BH}}$ column shows the Benjamini-Hochberg FDR for each of these genes.The columns med.x, ave.x and sd.x are the median, mean and standard deviation of 100 LTA's $x$  for each gene from 100 screening splits. Chrom refers to specific chromosome.}
\label{tab:RNAgene_table}
\medskip
\begin{tabular}{llllllll}
\hline
Gene& \multicolumn{1}{c}{Chrom} & \multicolumn{1}{c}{freq} & \multicolumn{1}{c}{FDR$_{\mbox{BH}}$} & \multicolumn{1}{c}{med(FDR)} & \multicolumn{1}{c}{med$(x)$} & \multicolumn{1}{c}{mean$(x)$} & \multicolumn{1}{c}{sd$(x)$}  \\
		\hline
  XIST & X &  91 & 1.78E-42 & 2.27E-19 & 8.92E-25 & 2.44E-23 & 7.24E-23 \\
  CYorf15A & Y &  91 & 1.03E-37 & 1.70E-16 & 1.64E-21 & 1.00E-20 & 2.58E-20 \\
    CYorf15B & Y &  91 & 1.31E-32 & 1.76E-14 & 1.92E-19 & 1.09E-18 & 2.13E-18 \\
    RPS4Y2 & Y &  91 & 1.31E-32 & 3.85E-14 & 2.32E-19 & 4.52E-18 & 1.49E-17 \\
    TTTY15 & Y &  91 & 1.53E-31 & 2.60E-13 & 1.31E-18 & 4.96E-17 & 2.03E-16 \\
  EIF1AY & Y &  91 & 8.78E-27 & 1.39E-11 & 3.89E-16 & 7.25E-15 & 2.11E-14 \\
   NLGN4Y & Y &  91 & 4.21E-24 & 7.37E-10 & 2.89E-14 & 3.86E-13 & 9.19E-13 \\
    UTY & Y &  91 & 7.01E-22 & 1.97E-09 & 1.63E-13 & 1.87E-12 & 5.62E-12 \\
  AC010889.1 & Y &  91 & 6.29E-21 & 4.93E-09 & 6.55E-13 & 3.18E-12 & 1.01E-11 \\
  RPS4Y1 & Y &  91 & 6.29E-21 & 1.16E-08 & 3.81E-13 & 2.03E-11 & 6.14E-11 \\
    KDM5D & Y &  91 & 1.82E-20 & 1.27E-08 & 1.41E-12 & 1.41E-11 & 4.08E-11 \\
    RP11-331F4.1 & 16 &  91 & 4.13E-20 & 2.12E-08 & 2.51E-12 & 2.82E-11 & 9.40E-11 \\
    DDX3Y & Y &  91 & 3.30E-18 & 2.16E-07 & 1.56E-11 & 6.59E-10 & 2.70E-09 \\
    NLGN4X & X &  91 & 2.24E-15 & 4.98E-06 & 7.24E-10 & 7.74E-09 & 1.86E-08 \\
    PNPLA4 & X &  89 & 3.65E-12 & 1.91E-04 & 9.08E-08 & 3.44E-07 & 8.81E-07 \\
   RP13-204A15.4 & X &  88 & 3.68E-13 & 2.41E-05 & 5.84E-09 & 1.84E-07 & 6.61E-07 \\
    AL137162.1 & 20 &  84 & 1.19E-10 & 4.20E-04 & 2.16E-07 & 1.48E-06 & 2.72E-06 \\
    RP5-1068H6.3 & 20 &  76 & 1.22E-09 & 9.72E-04 & 8.62E-07 & 2.95E-06 & 4.99E-06 \\
    AC016734.3 & 2 &  72 & 3.29E-09 & 7.49E-04 & 9.67E-07 & 1.10E-05 & 3.66E-05 \\
    RP11-143J12.1 & 18 &  64 & 1.47E-08 & 1.40E-03 & 2.47E-06 & 9.25E-06 & 1.58E-05 \\
    RPS4X & X &  64 & 1.56E-08 & 1.88E-03 & 3.49E-06 & 9.70E-06 & 1.69E-05 \\
    RP11-411G7.1 & 17 &  62 & 1.56E-08 & 1.80E-03 & 3.95E-06 & 1.14E-05 & 2.14E-05 \\
    RP11-863N1.1 & 18 &  61 & 3.35E-08 & 9.45E-04 & 2.68E-06 & 3.21E-05 & 8.07E-05 \\
    CTD-3116E22.2 & 19 &  58 & 3.56E-08 & 2.64E-03 & 6.68E-06 & 2.39E-05 & 6.72E-05 \\
    RP11-135F9.1 & 12 &  56 & 2.41E-08 & 2.73E-03 & 5.29E-06 & 1.87E-05 & 3.00E-05 \\
    RP11-624G17.1 & 11 &  56 & 3.59E-08 & 2.82E-03 & 5.76E-06 & 1.66E-05 & 3.06E-05 \\
    HDHD1 & X &  54 & 5.64E-08 & 1.58E-03 & 4.20E-06 & 3.77E-05 & 9.29E-05 \\
    RP11-21K20.1 & 12 &  48 & 5.64E-08 & 2.10E-03 & 7.33E-06 & 2.88E-05 & 4.67E-05 \\
 \hline
\end{tabular}
\end{minipage}
\end{table}

Of the 69 individuals, there are 40 females and 29 males. From the whole group, 20 female and 15 male subjects were randomly selected to construct the modeling split, (the remaining subjects formed the screening split). Assuming the data follows a negative binomial distribution \cite{anders2010diff}, the p-value for each gene was calculated using a generalized likelihood ratio test comparing male and female subjects in the modeling split. The modeling data consists of these 17310 p-values. Similar p-values from the screening split provided the screening data.

A mixture of a $Uniform$ and a $Beta$ distribution was fitted to the modeling data p-values and was adjusted to obtain the empirical fit as described in Subsection \ref{subs_emp_fit}. Figure~\ref{fig:whole_pvalue} illustrates the effect of the adjustment on the whole data. An initial mixture model fit $f(x)=0.993f_0^*(x) + 0.007f_1^*(x)$ with  the $Uniform(0,1)$ density $f_0^*(x)$, and the $Beta(\alpha=0.064,\beta=1.517)$ density $f_1^*(x)$ is illustrated in Panel~\ref{fig: whole1}, while the adjusted mixture model $f(x)=0.994\hat{f}_0(x) + 0.006\hat{f}_1(x)$, according to Equation \eqref{eq:fit}, is featured in Panel~\ref{fig: whole2}.

For the analysis, the adjusted fit from a modeling data was used to calculate the tail-area Fdr associated with the p-value in the corresponding screening data for each gene. Genes with Fdr less than 0.01 were detected as significantly differentially expressed between male and female subjects. The process was repeated 100 times.

100 sample splits and subsequent screening detected a total of 83 significant genes out of 17310.  Of these 83 detected significant genes, 56 appeared more than once in the 100 repetitions of the splitting and screening. Table~\ref{tab:RNAgene_table} presents a partial summary of the detection results that includes 28 most frequently detected significant genes with tail-area Fdr$<0.01$ (detection frequencies 40 or higher). The ``freq" column indicates the frequency of occurrence for the corresponding gene in the 100 screening splits. Note in particular, the top-ranked gene XIST (X inactive specific transcript) is expressed only in females to silence one of the two X chromosomes; this provides the dosage equivalence between females and males. The other most frequently detected significant genes appear on either X or Y chromosomes, which is expected since they are differentially expressed between males and females.

\begin{figure}[!htpb]
\centering
\begin{subfigure}[b]{0.4\textwidth}
%\vskip -.2in
\includegraphics[width=\textwidth]{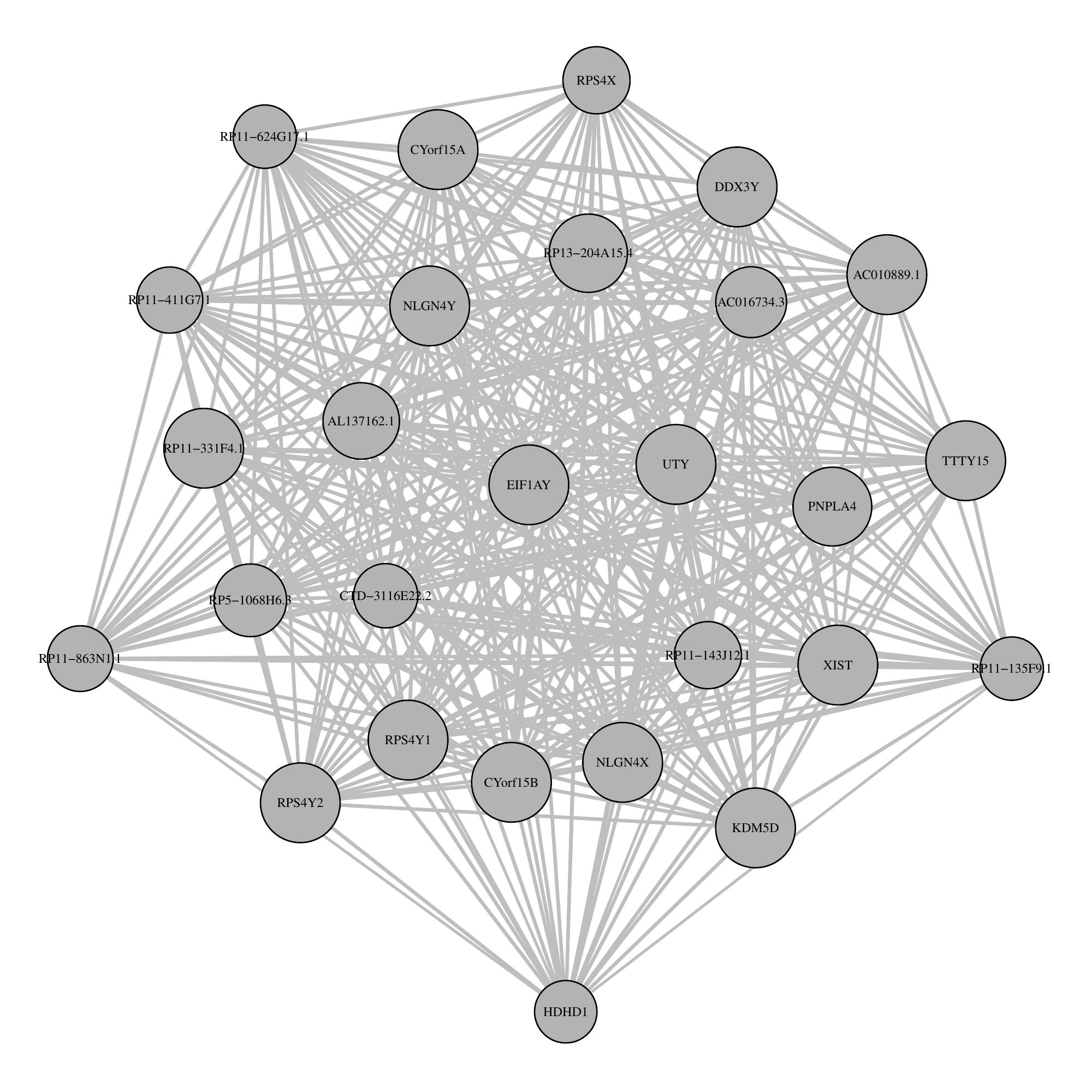}
\caption{F-association for 27 most significant genes}
\label{fig: sub-network3}
\end{subfigure}
%\hfill
\begin{subfigure}[b]{0.4\textwidth}
%\vskip -.2in
\includegraphics[width=\textwidth]{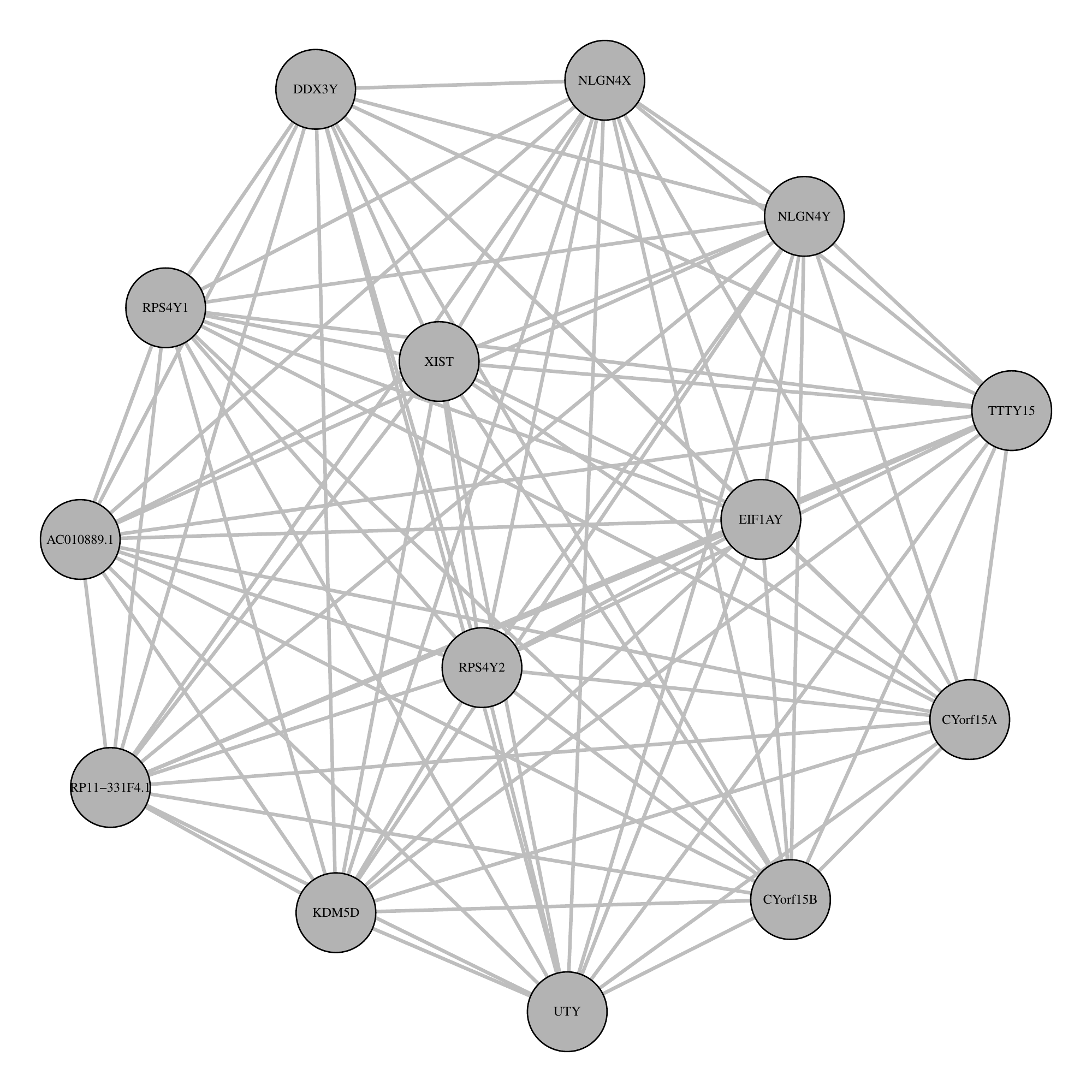}
\caption{F-association for 14 most significant genes}
\label{fig: sub-network4}
\end{subfigure}
\medskip
\caption{F-association for the most significant genes appearing more than 50 times in (a) and more than 90 times in part (b). The Fdr threshold is at 0.01.}
\label{fig: RNAnetwork2}
\end{figure}

Figure \ref{fig: RNAnetwork2} shows the F-association plot for the most significant genes. Since we are using p-values for the analysis, detected genes cannot be flagged as over-expressed or under-expressed (unlike LTA values), but can only be detected as significantly differentially expressed in females compared to males. Therefore the color scheme of Figure \ref{fig: network} is absent in Figure \ref{fig: RNAnetwork2}.  Each node represents a gene, and the gray link between nodes indicates the pair of genes that are simultaneously differentially expressed. In Figure~\ref{fig: sub-network3}, most of the inner clustered genes come from the top-ranked genes from Table~\ref{tab:RNAgene_table}, and the more outward the nodes are in the frequency network, the lower the frequency of being differentially expressed.

\subsection{Simulation Studies}
\label{subs:simulation}
In this section, we present several simulation studies featuring the proposed screening. Subsection \ref{subs:simulation1} includes a continuous data analysis and detailed power and precision comparison with Efron's whole data fit method. Subsection \ref{subs:simulation2} presents the specificity of the method for a simulated data without any signal. Subsection \ref{subs:simulation3} illustrates the analysis of a discrete data, whereas in subsection \ref{subs:simulation4} we analyze a similar discrete model simulation with a larger proportion of the signal.

\subsubsection{Continuous Data Simulation}
\label{subs:simulation1}

We simulated a data set with a treatment group of 50 subjects and a control group of 50 subjects where 1000 expressions were simulated for each subject. Out of the total 1000  variables, the first 30 were set to be the ``non-null" cases (expressions were simulated from distributions different for the treatment and control groups); while the last 970 variables were set to be  the ``null" cases (expressions were simulated from the same distribution for the treatment and control groups).

{\renewcommand{\arraystretch}{3}
\begin{table}[!h]
%\vspace{-0.3 cm}
\begin{minipage}[c]{0.95\linewidth}
\centering
\caption{The simulation parameters for 30 non-null variables. Here 10 values for $\mu_{c_1}$ were generated from an $N(5,1)$ distribution, one for each of the variables $11$ to $20$ in the treatment group. Similarly, 10 values for $\mu_{c_2}$ were generated from an $N(7,1)$ distribution, one for each of the output variables $21$ to $30$ in the treatment group.}
\label{table_sim_par}
\medskip
\begin{tabular}{ccc*2{>{\renewcommand{\arraystretch}{1}}c}}
\hline
{\multirow{2}{*}{Output Variables}} &\multicolumn{2}{c}{Mean}&{\multirow{2}{*}{\shortstack{Variance \\ (treatment and control)}}}\\
\cline{2-3}&treatment&\ control\ &\\
\hlineB{3}
$\{1,2\}$ & $(7,7)'$ & $(6,6)'$ & $2*\begin{bmatrix}
1   & 0.8 \\
0.8 &  1  \\
\end{bmatrix}$ \\ [4 pt]
\hline
$\{3,4\}$ & $(5,5)'$ & $(6,6)' $ & $2*\begin{bmatrix}
1   & -0.8 \\
-0.8 &  1  \\
\end{bmatrix}$\\ [4 pt]
\hline
$\{5,6,7\}$ & $(7.5,7.5,7.5)'$ & $(6,6,6)' $ &$2*
\begin{bmatrix}
1   & 0.75 & 0.8 \\
0.75 & 1 & 0.9 \\
0.8 & 0.9 & 1  \\
\end{bmatrix}$\\ [10 pt]
\hline
$\{8,9,10\}$ & $(4.5,4.5,4.5)'$ & $(6,6,6)' $ &$2*
\begin{bmatrix}
1   & -0.85 & -0.9 \\
-.85 & 1 & 0.61 \\
-0.9 & 0.61 & 1  \\
\end{bmatrix}$\\ [10 pt]
\hline
$\{11,12,\ldots,20\}$ (independent)& $\mu_{c_1}$&6&2\\
\hline
$\{21,22,\ldots,30\}$ (independent)& $\mu_{c_2}$&6&2\\
\hline
\end{tabular}
\end{minipage}
\vspace{-0.3 cm}
\end{table}
}

Further, to show that the F-association plots constructed using the detection frequencies can pick up existing inter-relationships between screened cases, we added a correlation structure among the first 10 non-null variables. For a gene expression study, if some genes are inter-related with each other, they will work in tandem in any subject no matter whether from the control or from the treatment group, although their expression levels can be different between the two groups. Keeping that in mind, for the simulated data the same correlation structures were applied to both the treatment and the control group while keeping the non-null means different in the two groups.

We used the normal distribution for the simulation. The $N(6,2)$ distribution was used for all 970 null variables for each subject in the treatment and the control group. The treatment and control group parameters used to simulate the 30 non-null variables are described in Table~\ref{table_sim_par}, where the mean and the variance columns illustrate the Normal distribution mean vector and variance-covariance matrices used for simulation. Variables $\{1,2\}$, $\{3,4\}$, $\{5,6,7\}$ and $\{8,9,10\}$  are clustered together with the covariance structure shown in the table.  Whereas, variables 11 to 30 were simulated independently from $N(\mu_{c_1},6)$. 10 values for $\mu_{c_1}$ were generated from an $N(5,1)$ distribution, one for the mean of each of the variables $11$ to $20$ in the treatment group. Similarly, 10 values for $\mu_{c_2}$ were generated from an $N(7,1)$ distribution, one for the mean of each of the output variables $21$ to $30$ in the treatment group which are independently simulated from $N(\mu_{c_2}, 6)$ distributions.

\begin{figure}[!h]
\centering
\begin{subfigure}{.5\textwidth}
  \centering
  \includegraphics[width=.9\linewidth]{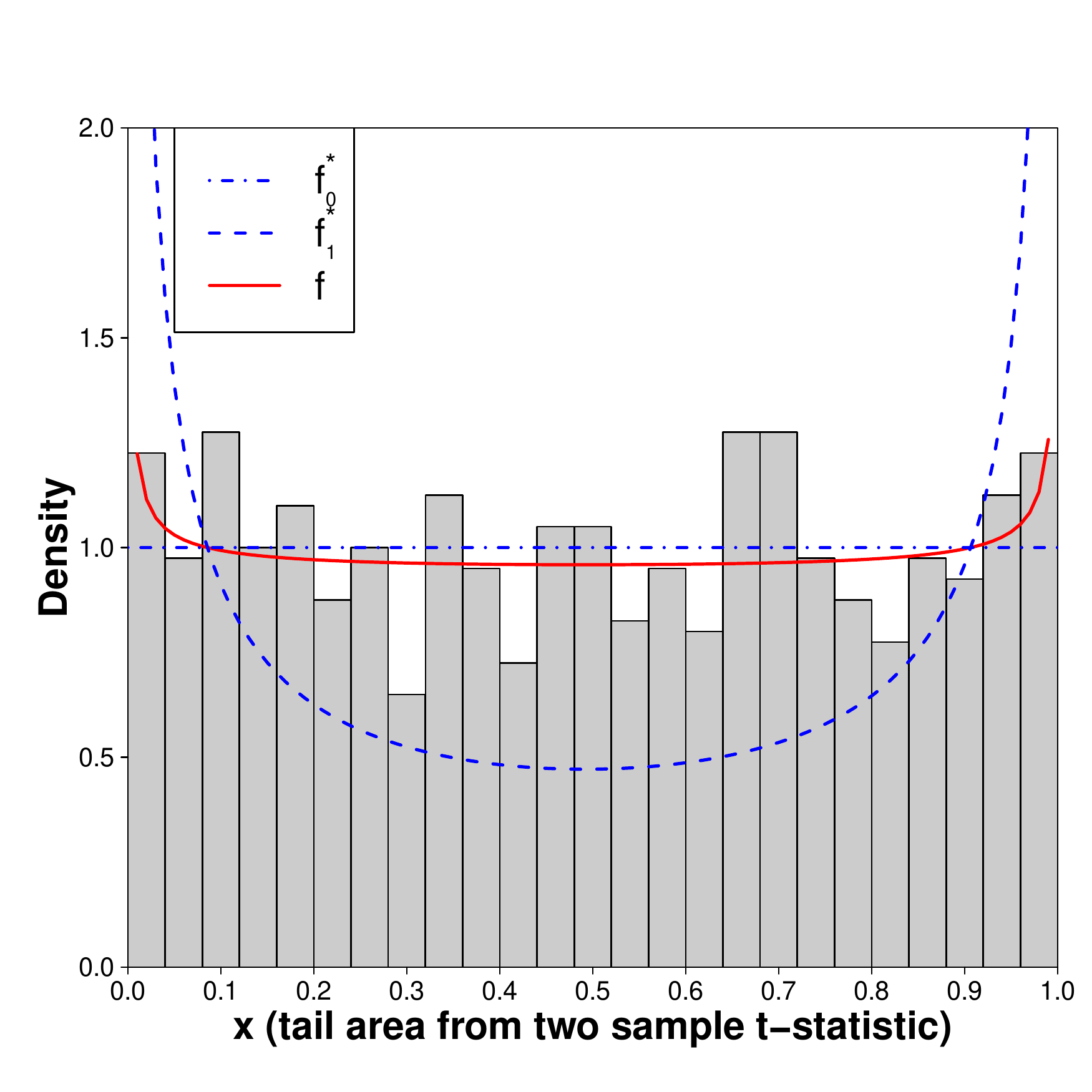}
  \caption{Uniform-Beta mixture distribution.}
  \label{fig: sim_mix1}
\end{subfigure}%
\begin{subfigure}{.5\textwidth}
  \centering
  \includegraphics[width=.9\linewidth]{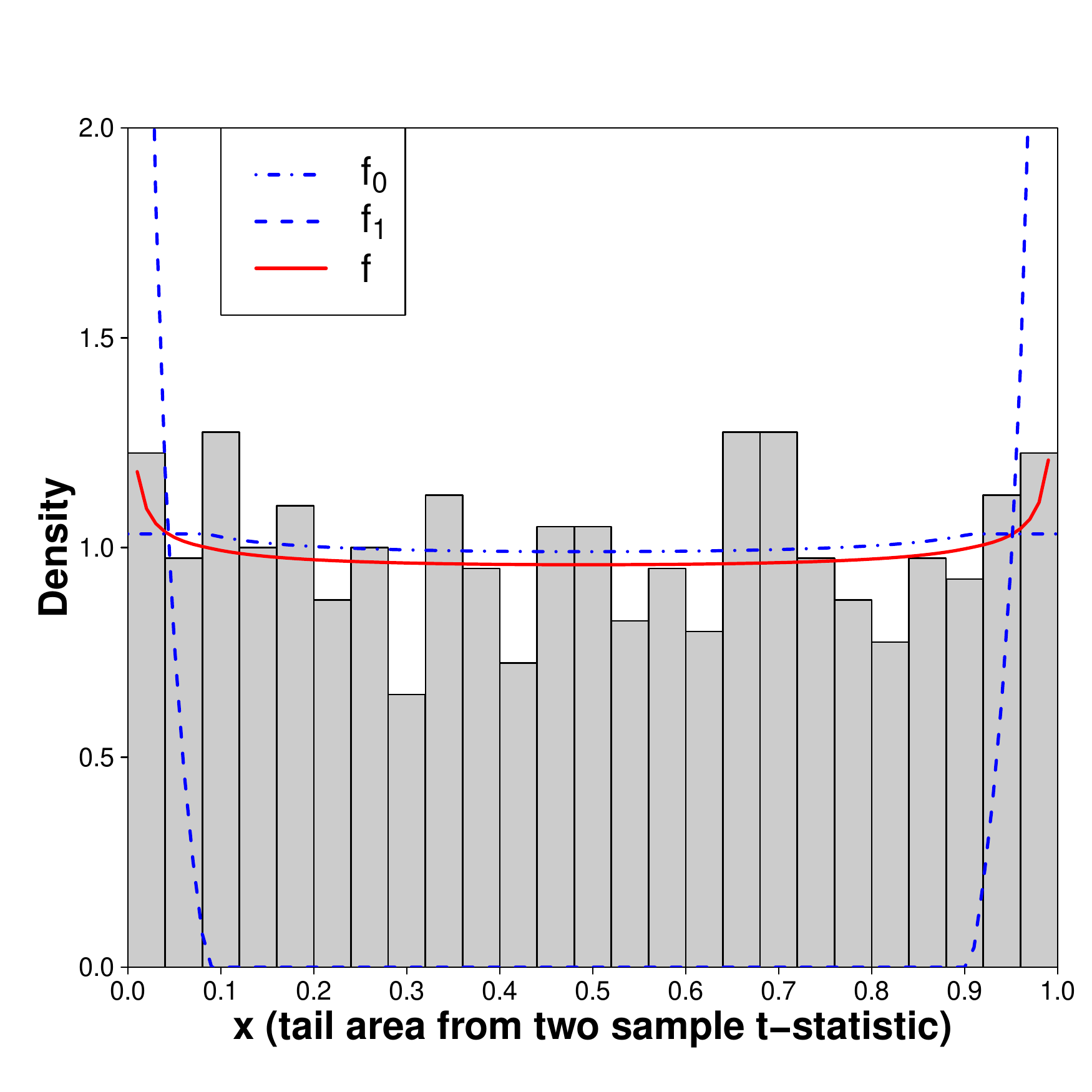}
  \caption{Adjusted Uniform-Beta mixture distribution.}
  \label{fig: sim_mix2}
\end{subfigure}
\medskip
\caption{The histogram of $x_i$ (left-tail-area for the observed two sample t-statistic for the simulated variables $i=1,2,\ldots,1000$) from a screening data set consisting of half of the control and the treatment groups respectively. Superimposed on Figure~\ref{fig: sim_mix1} are the fitted Uniform, Beta and the  mixture distribution obtained from the corresponding modeling split as $\hat{f}(x)=0.923f_{0}^{\ast}(x)+0.077f_{1}^{\ast}(x)$ where $f_{0}^{\ast}$ is the $Uniform(0,1)$ pdf and $f_{1}^{\ast}$ is the $Beta(0.341,0.319)$ pdf. The red solid line shows the fitted mixture distribution. Figure~\ref{fig: sim_mix2} shows the empirical null fit adjusted from the fitted Uniform-Beta mixture distribution to $\hat{f}(x)=0.974\hat{f}_{0}(x)+0.026\hat{f}_{1}(x)$ as in Equation \eqref{eq:fit}.}
\label{fig: sim_mix}
\end{figure}

The 100 subjects were randomly split into a modeling and a screening set, each consisting of 25 subjects from the treatment group and 25 subjects from the control group. The data points were defined as $x_i=P(t<t_i)$, where $t_i$ is the two-sample t-test statistic for each output variable from the 25 control and the 25 treatment samples in the modeling split. Then a mixture of a Uniform and a Beta distribution was fitted to the modeling split $x_i$'s and was adjusted to better capture the background and the signal similar to \eqref{eq:fit} in Section \ref{subs:microarray}. %Figures \ref{fig: sim_mix1} and \ref{fig: sim_mix2} show a fit from one particular modeling split.

Figure~\ref{fig: sim_mix} shows the histogram of $x_i$ (left-tail-area for the observed two sample t-statistic for the simulated variables $i=1,2,\ldots,1000$) from a particular screening data. Figure~\ref{fig: sim_mix1} illustrates the initial mixture model $\hat{f}(x)=0.923f_{0}^{\ast}(x)+0.077f_{1}^{\ast}(x)$, where $f_{0}^{\ast}$ is the $Uniform(0,1)$ pdf and $f_{1}^{\ast}$ is the $Beta(0.341,0.319)$ pdf, fitted from the modeling data. The red solid line shows the fitted mixture distribution. Figure~\ref{fig: sim_mix2} presents the adjusted empirical null fit $\hat{f}(x)=0.974\hat{f}_{0}(x)+0.026\hat{f}_{1}(x)$ according to Equation \eqref{eq:fit}.

\begin{figure}[!h]
\centering
\vspace{-1.5 cm}
\begin{subfigure}{.8\textwidth}
  \centering
  \includegraphics[width=.95\linewidth]{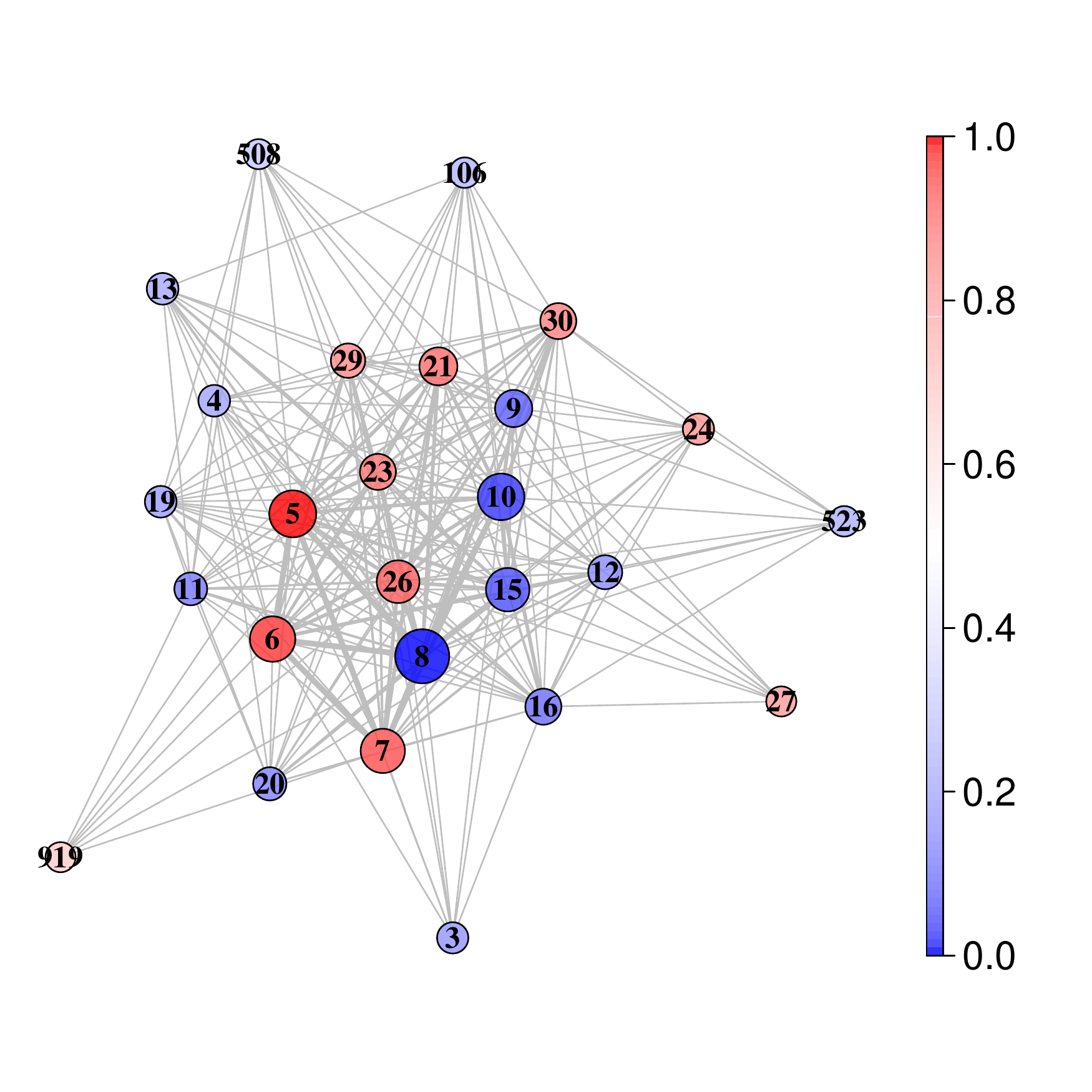}
  \vspace{-1.3 cm}
  \caption{F-association for 26 most significant variables.}
  \label{fig: network1}
\end{subfigure}%
\par
\vspace{-1.2 cm}
\begin{subfigure}{.9\textwidth}
  \centering
  \includegraphics[width=0.9\linewidth]{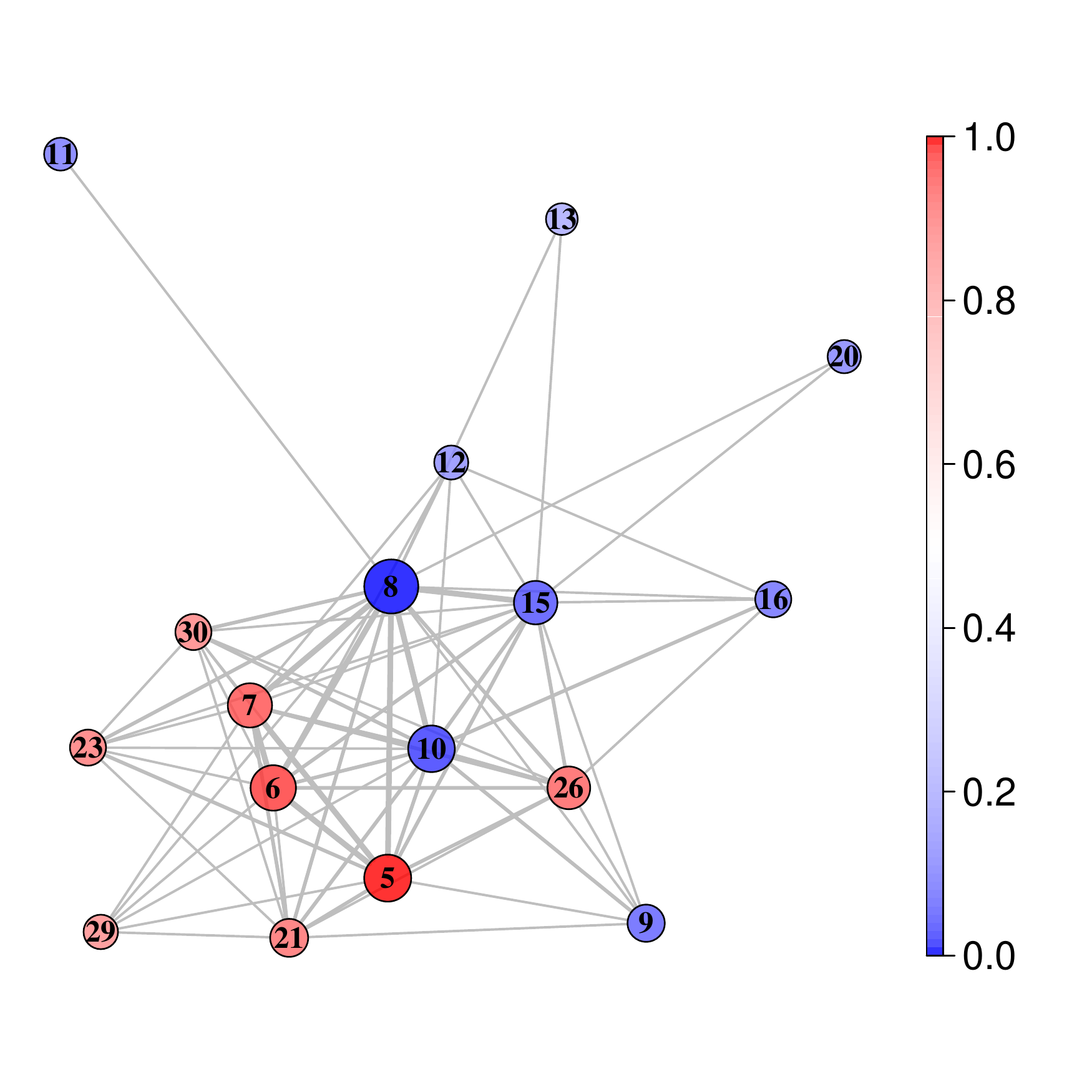}
  \vspace{-1.5 cm}
  \caption{Simplified F-association for \ref{fig: network1}.}
  \label{fig: network2}
\end{subfigure}
\medskip
\caption{Figure~\ref{fig: network1} is the F-association plot of 26 significant variables appearing at least twice in the 100 sample splits by using the tail-area Fdr$\le 0.1$. Figure~\ref{fig: network2} is the simplified network of Figure~\ref{fig: network1} by deleting variables with less than 5 connected edges. Node color blue indicates the gene was expressed at a lower level in the treatment group compared to the control group, whereas, red implies the gene was expressed at a higher level in the treatment group. }
\label{fig: sim_network}
\vspace{-1.5 cm}
\end{figure}

\begin{figure}[!h]
\vspace{-0.5 cm}
\hspace{-0.8 cm}
\centering
\begin{subfigure}{.52\textwidth}
  \centering
  \includegraphics[width=.95\linewidth]{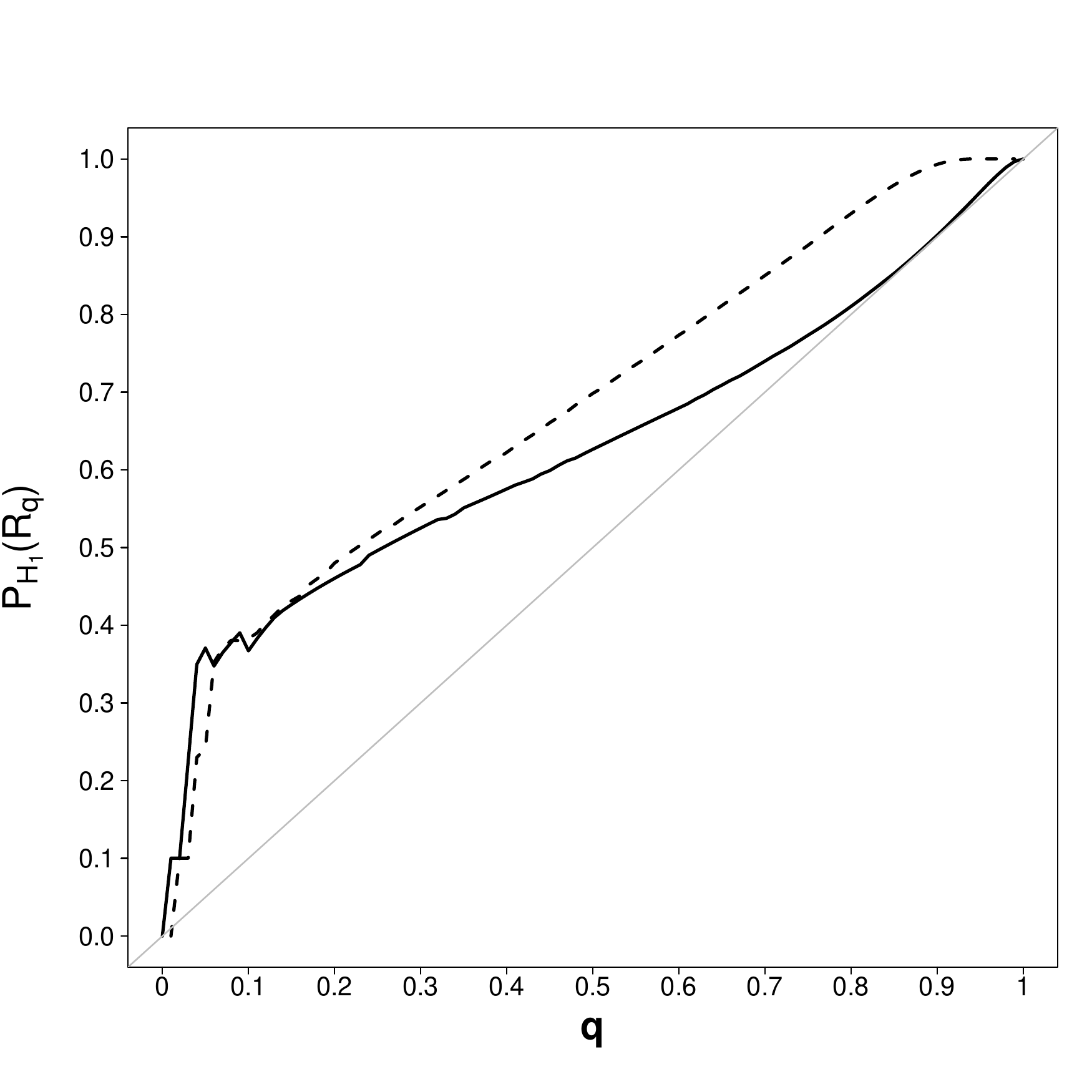}
  \caption{Power curve comparison. }
  \label{fig: recall_lfdr}
\end{subfigure}%
\begin{subfigure}{.52\textwidth}
  \centering
  \includegraphics[width=.95\linewidth]{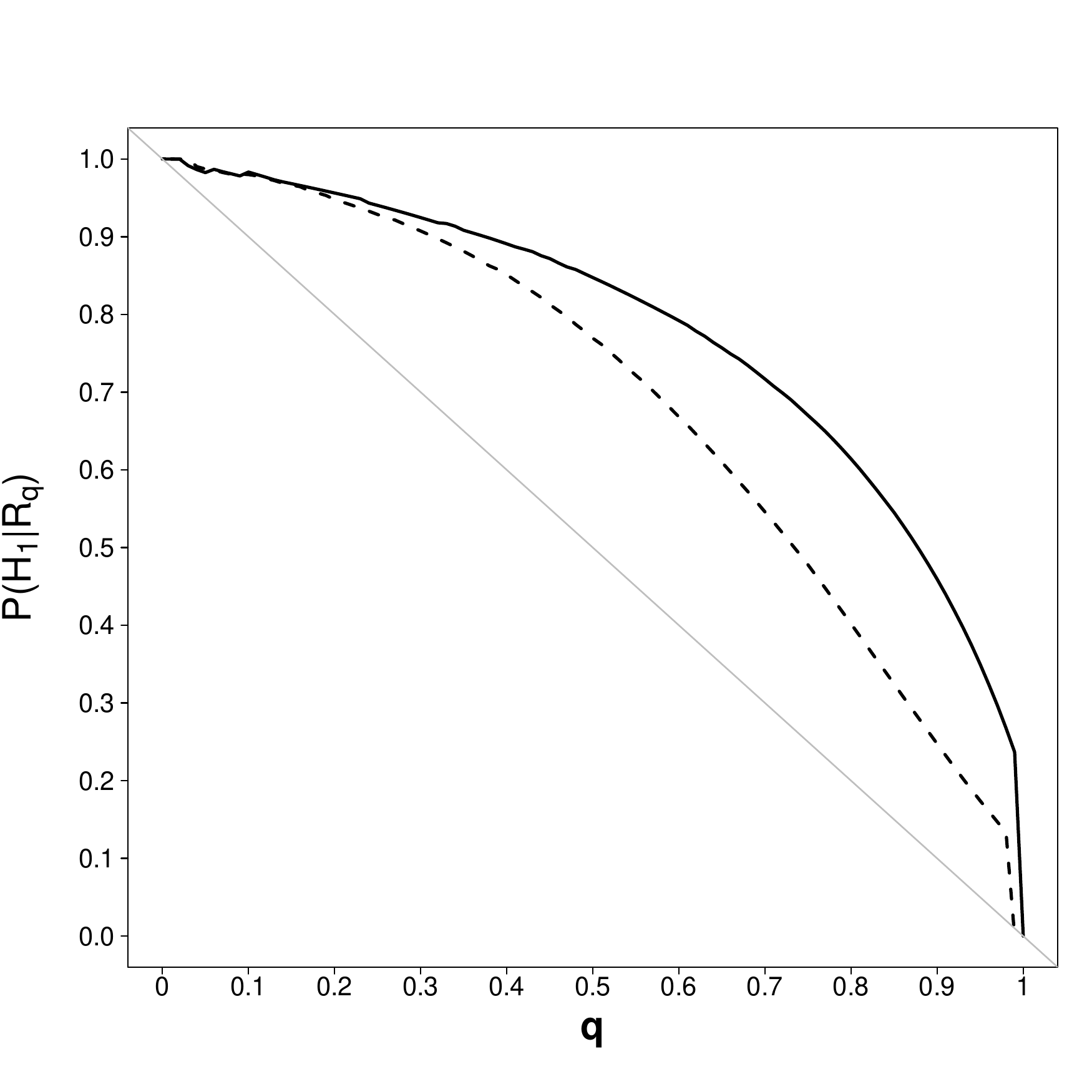}
  \caption{Precision curve comparison.}
  \label{fig: precision_lfdr}
\end{subfigure}
\medskip
\caption{Figure~\ref{fig: recall_lfdr} is the power curve comparison and Figure~\ref{fig: precision_lfdr} is the precision curve comparison between the whole-data fit/screening method and the proposed sample splitting method. Here $q$ is the cutoff point of \textbf{the local fdr}. The solid line represents power or precision using the whole data fit/screening method, and the dashed line represents the same using the proposed method with 100 sample splits.}
\label{fig: power_comp_lfdr}
\end{figure}

Here we used  tail-area Fdr screening to construct the frequency table (Table~\ref{table_sim_freq1}) and the F-association plots (Figure \ref{fig: sim_network}). The local fdr screening results are used for comparison purposes in Figure \ref{fig: power_comp_lfdr}.

The two-sample $t$  for each variable in the screening split and it's left-tail-area $x_i$ formed the screening data. The  modeling split fit \eqref{eq:fit} was used to obtain the tail-area Fdr associated with each screening data point $x_i$. The variables with tail-area Fdr less than $0.1$ were detected as significant. The process was repeated 100 times.  Table~\ref{table_sim_freq1} shows 26 most frequently detected significant variables from 100 sample splits with the tail-area Fdr$<0.1$ (detection frequencies 2 or higher). The ``freq" column shows the frequency of detection for the corresponding variable in 100 screening splits.

\begin{table}[!h]
\begin{minipage}[c]{0.95\linewidth}
%\vspace{-0.1 cm}
\centering
\caption{Screening result for simulated continuous data where variables 1 to 30 are set to be non-null. Table shows 26 detected significant variables from 100 sample splits with the tail-area Fdr$<0.1$ (detection frequencies 2 or higher). The ``freq" column shows the frequency of detection for the corresponding variable ($i$) in 100 screening splits. The values in the third column show the median tail-area Fdr from 100 sample splits for each gene. The columns med.x, ave.x and sd.x are the median, mean and standard deviation of the LTA's $x$  for each variable computed from 100 randomly chosen screening data sets.}
\label{table_sim_freq1}
\medskip
\begin{tabular}{crrrrc}
\hline
\textbf{variable ($i$)} & \multicolumn{1}{c}{\textbf{freq}} & \multicolumn{1}{c}{\textbf{med(Fdr)}} & \multicolumn{1}{c}{\textbf{med}$(x)$} & \multicolumn{1}{c}{\textbf{avg}$(x)$} & \multicolumn{1}{c}{\textbf{sd}$(x)$} \\
		\hline
 8 &     63 & 0.03660 & 0.00006 & 0.00058 & 0.00153 \\
 5 &     45 & 0.04460 & 0.99987 & 0.99928 & 0.00146 \\
 10 &     44 & 0.03740 & 0.00040 & 0.00266 & 0.00862 \\
  6 &     41 & 0.04690 & 0.99979 & 0.99898 & 0.00216 \\
  7 &     38 & 0.05290 & 0.99971 & 0.99868 & 0.00303 \\
  15 &     36 & 0.04490 & 0.00051 & 0.00399 & 0.01051 \\
  26 &     34 & 0.05280 & 0.99962 & 0.99826 & 0.00362 \\
  21 &     22 & 0.05030 & 0.99923 & 0.99583 & 0.01447 \\
  9 &     20 & 0.04850 & 0.00213 & 0.01042 & 0.01890 \\
  16 &     17 & 0.04820 & 0.00233 & 0.01138 & 0.02408 \\
  23 &     17 & 0.05830 & 0.99841 & 0.99473 & 0.01289 \\
  30 &     17 & 0.07370 & 0.99819 & 0.99265 & 0.01572 \\
  29 &     13 & 0.04630 & 0.99760 & 0.99270 & 0.01855 \\
  12 &     12 & 0.04980 & 0.00648 & 0.01892 & 0.03579 \\
  20 &     10 & 0.06330 & 0.00518 & 0.02024 & 0.03748 \\
  11 &      9 & 0.06040 & 0.00311 & 0.01547 & 0.03765 \\
  13 &      6 & 0.06720 & 0.01671 & 0.02783 & 0.03881 \\
  19 &      6 & 0.06930 & 0.00958 & 0.01910 & 0.02499 \\
   4 &      6 & 0.06720 & 0.01157 & 0.02577 & 0.04406 \\
  24 &      5 & 0.04710 & 0.99692 & 0.99130 & 0.01475 \\
  3 &      5 & 0.06420 & 0.00861 & 0.02270 & 0.03694 \\
  106 &      3 & 0.03140 & 0.02360 & 0.04740 & 0.06810 \\
  523 &      3 & 0.04050 & 0.02281 & 0.05717 & 0.10463 \\
  27 &      2 & 0.07010 & 0.98869 & 0.96776 & 0.05341 \\
  508 &      2 & 0.07170 & 0.04119 & 0.07933 & 0.10046 \\
  919 &      2 & 0.05420 & 0.92189 & 0.89232 & 0.12438 \\
\hline
\end{tabular}
\end{minipage}
\end{table}

Note that, although variables 1 to 30 out of the 1000 simulated variables were set as non-null, the groups of variables 5, 6, 7 and 8, 9, 10 deviated the most from the null. The mean vectors for variables 1, 2 and 3, 4 did not deviate enough from the null mean to produce significantly large or small t-statistic values. The analysis was done using the t-statistic tail-area and not the original normal distribution, thus naturally non-null variables 1, 2 were not captured in the screening process.

Figure~\ref{fig: network1} presents the F-association plot of 26 significant variables appearing at least twice in the 100 sample splits by using the tail-area Fdr$\le 0.1$. Figure~\ref{fig: network2} is the simplified network of Figure~\ref{fig: network1}, obtained by deleting variables with less than 5 connected edges. Here node color blue indicates the gene was expressed at a lower level in the treatment group compared to the control group, whereas, red implies the gene was expressed at a higher level in the treatment group.

The F-association plots in Figures \ref{fig: network1} and \ref{fig: network2} show that among the detected significant variables, the groups with strongest correlation structure, one with variables 5, 6 and 7 and another with variables 8, 10 were captured successfully. However, the plot did not capture all correlated genes (e.g., gene 9 that had a low correlation with gene 10 but had a higher correlation with gene 8 did not show up as in the group with $\{8, 10\}$).

The association featured in the network plot depends on the difference between control and treatment groups in the observed data, and while the inherent dependence structure might drive the difference, the difference itself might not always indicate a dependence. Hence the plot should be used for exploration and not to prove a causal association. A more detailed discussion about the limitation of the F-network is included in Section \ref{s_discussion}, while a justification for the association is presented in Appendix \ref{app:F_association}.\\

\textbf{\underline{Power and precision comparison with the whole data fit method.}}
%\medskip\\

To compare the efficacy of the proposed method with the whole data fit/screening, we present the power (recall) and precision (Equations \eqref{recall} and  \eqref{precision}) comparisons in Figure \ref{fig: power_comp_lfdr}. Since Efron \cite{Efron07} used the local fdr for screening, for comparison purposes with repeated sample splitting, the combined rejection region based on local fdr screening as in \eqref{R_comb2} is used in Figure \ref{fig: power_comp_lfdr} along with the rejection region from the whole data fit and the local fdr screening method.

The combined rejection region from repeated sample splits results in a larger rejection region hence providing higher power as evident in Figure \ref{fig: recall_lfdr}. But the inclusion of all detections increases the number of false discoveries and consequently decreases the  precision as seen in Figure \ref{fig: precision_lfdr}. Note that, even when the entire $R(q)$ was used with the proposed method as the rejection region,  at relevant $q$ values 0.1 to 0.3 (reasonable cutoff points for local fdr) the sample splitting technique shows a level of precision that is on par with the whole data fit/screening method.

However, we are proposing that only the variables with high detection frequencies should be screened as potential true discoveries and not the entire $R(q)$. By adding a high frequency criterion (as in \eqref{eq:freq_cutoff_N}) to the combined rejection region, the precision can be set at any desired fixed level for all $q$. Whereas, with enough repeated sample splitting, the set of variables in $R(q)$ with high frequencies is unlikely to get much smaller compared to the rejection set produced by the existing methods, with sufficiently large repetition of sample splitting, the high frequency screening can be set to achieve intended level of precision without significant loss of power.

For example, 21 of 26 detections (about 81\%) in Table~\ref{table_sim_freq1} with detection frequencies of 2 or higher were true discoveries. If  variables with detection frequency 5 or higher are considered, all 21 detections are true discoveries (100\%).\\

\textbf{\underline{Critical detection frequency and $Fdr_o$.}}

For a fixed $q$, we screen significant genes at every splitting/screening step if the tail-area Fdr$<q$ in the verification split. This leads to the rejection region from an individual split according to Equation \eqref{Rq}, which can be calculated by a numerical grid search for $x\in [0,1]$, so that Equation \eqref{Rq} holds. The overall rejection region $R(q)$ is constructed by combining rejection regions from 100 repeated splits, according to equation \eqref{R_comb1}.

 For example, for the simulated data generated with parameters in Table~\ref{table_sim_par}, if we keep $q=0.1$, the combined rejection regions from 100 repeated random sample split/screening is $R(q)=\{x: x\in[0, 0.0001] \bigcup[0.9999, 1]\}$.

 The $Fdr_o$ and critical detection calculation steps described in Subsection \ref{subs:power}, Equations \eqref{Fdr_o} and \eqref{eq:freq_cutoff_N} use a fitted mixture distribution on the entire data without sample splitting (as described in step (iii) under the heading \textbf{``Rejection Region and Power Calculation Steps"} in that subsection). For this particular simulated data set,  the whole data Uniform/Beta mixture fit on left-tail areas $x$  is given by $f(x) = 0.937\times Uniform + 0.063\times Beta(0.19,0.18)$, which in its tail-adjusted form can be written as $f(x) = 0.969\times\tilde{f}_0(x) + 0.031\times\tilde{f}_1(x)$  (adjusted according to Equation (3.2)).

When the splitting/screening in repeated $N=100$ times, to keep $Fdr_o<q^*$, the critical detection frequency for potential true significance can be calculated from Equation \eqref{eq:freq_cutoff_N} as
\[N\times \dfrac{\tilde{p}_0\times\int_{R(q)}\tilde{f}_0(x)dx}{q^*}=  100\times\dfrac{0.969\times \int\limits_{[0, 0.0001] \bigcup[0.9999, 1]}\tilde{f}_0(x)dx}{q^*},\]
where $q^*$ can be user-defined such as 0.05, 0.1, and 0.2, etc. A numerical integration procedure is used for this part of the calculation in our R-package.

To generalize the concept for any $N$, one might just consider the critical detection ``relative" frequency by
\[\dfrac{\tilde{p}_0\times\int_{R(q)}\tilde{f}_0(x)dx}{q^*}=  \dfrac{0.969\times \int\limits_{[0, 0.0001] \bigcup[0.9999, 1]}\tilde{f}_0(x)dx}{q^*}.\]

\begin{figure}[!h]
\vspace{-0.8 cm}
\hspace{-1 cm}
\centering
\begin{subfigure}{.52\textwidth}
  \centering
  \includegraphics[width=.97\linewidth]{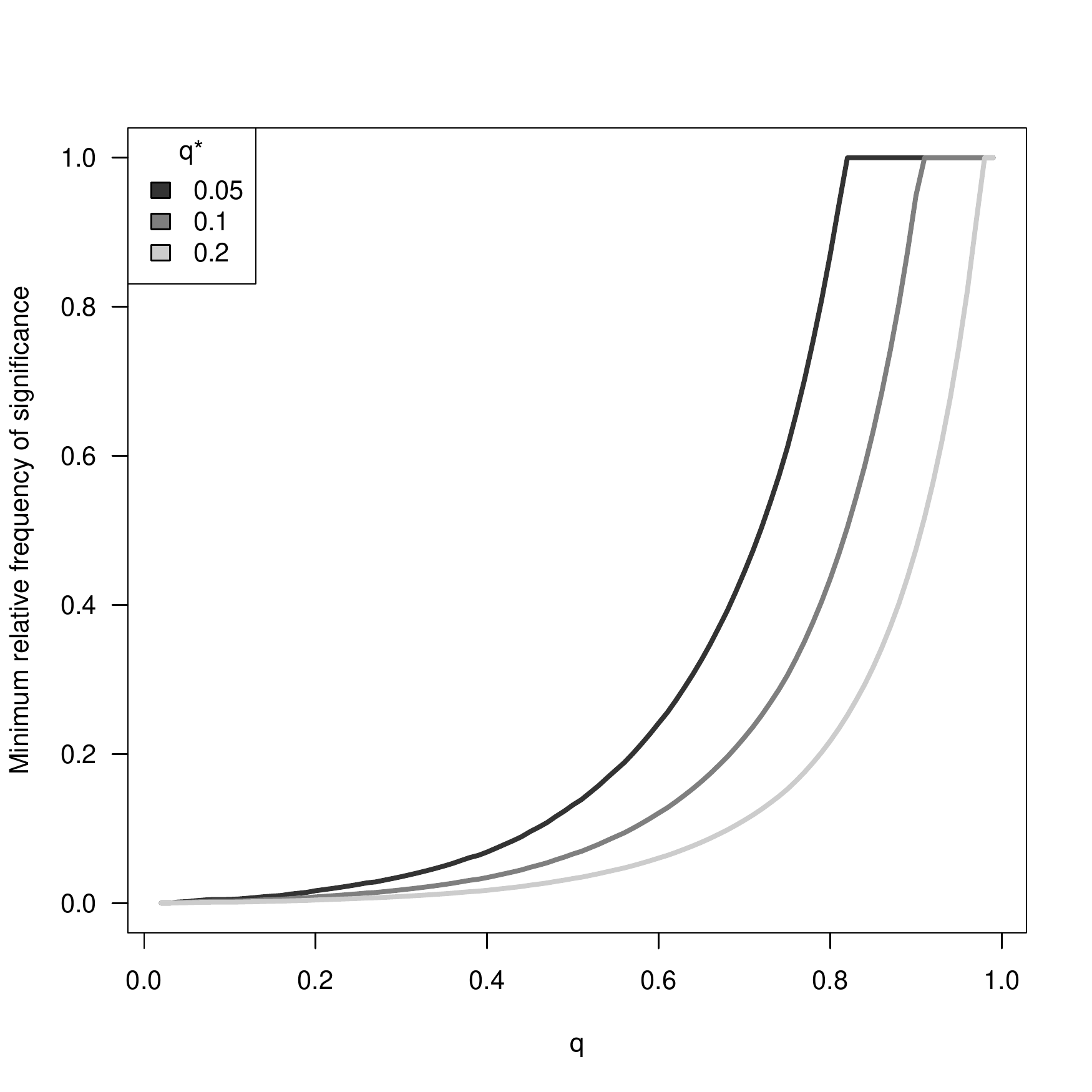}
  \caption{The tail-$Fdr$  cutoff point $q$ used for each screening vs. relative frequency of detection needed to attain $Fdr_o<q^*$.}
  \label{fig: Fdr_O1}
\end{subfigure}%
\begin{subfigure}{.54\textwidth}
  \centering
  \includegraphics[width=.97\linewidth]{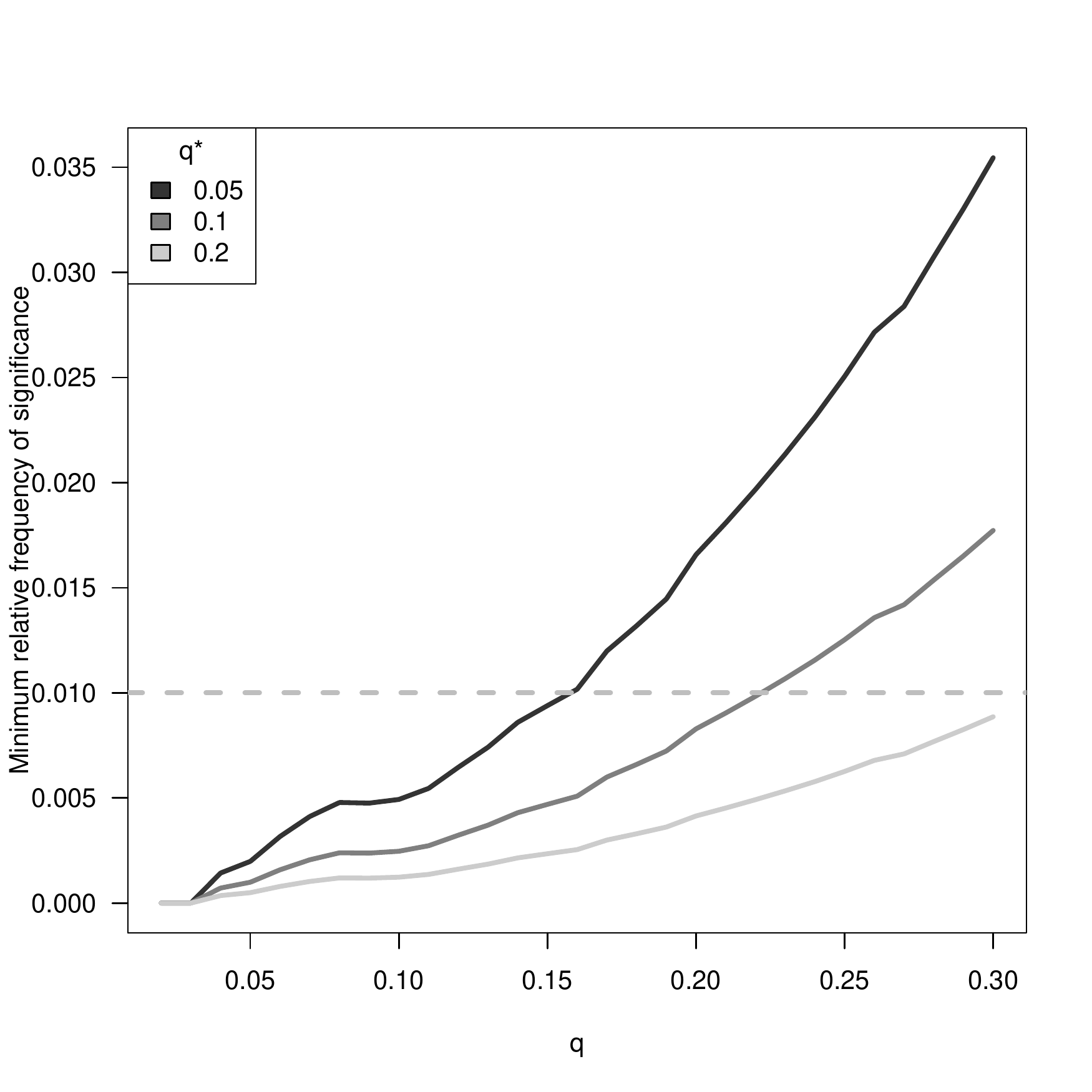}
  \caption{Zoomed-in section of plot \ref{fig: Fdr_O1} }
  \label{fig: Fdr_O2}
\end{subfigure}
\medskip
\caption{
Critical relative frequency of detection needed to achieve $Fdr_o\leq q^*$ according to Equation \eqref{eq:freq_cutoff}, when the criterion used to detect potential significant cases from each individual screening split is ``tail-area Fdr$<q$".}
\label{fig: FdrO}
%\vspace{-0.5 cm}
\end{figure}

Figure \ref{fig: FdrO} illustrates the critical relative frequency of detection required to keep the overall combined false discovery rate $Fdr_o$ bounded at $q^*$ = $0.2$, $0.1$ or $0.05$, when each screening-split uses detection criterion \textit{tail-area Fdr} $< q$.

For example, in  Figure \ref{fig: Fdr_O2}, at $q=0.15$ the graph for $q^*=0.05$ shows the minimum relative frequency needed is about .01. This implies that for each screening-split, if the detection criterion employed is ``a variable is  significant when its corresponding tail-area Fdr$<0.15$" and the process is repeated 500 times, any variable needs to be detected more than $(500\times 0.01)=5$ times to be identified as a potential true discovery in order to keep $Fdr_o<0.05$ for the combined screening process. If the process is repeated 1000 times, any variable needs to be detected more than 10 times to be identified as a potential true discovery in order to keep $Fdr_o<0.05$ for the combined screening process.

\begin{figure}[!htb]
\centering
\begin{subfigure}{.5\textwidth}
  \centering
  \includegraphics[width=.9\linewidth]{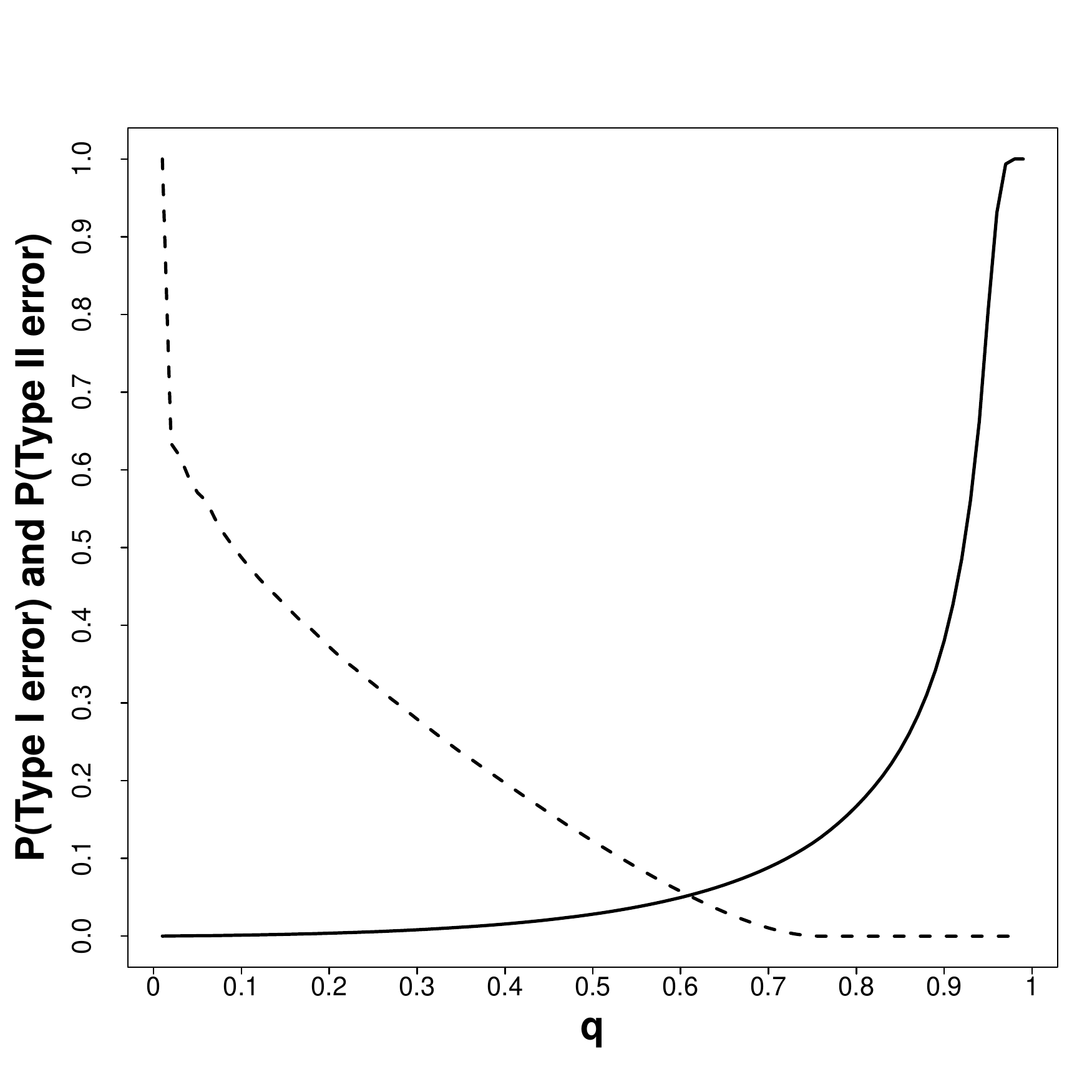}
  \caption{Type I and II errors.}
  \label{fig: uncond_type}
\end{subfigure}%
\begin{subfigure}{.5\textwidth}
  \centering
  \includegraphics[width=.9\linewidth]{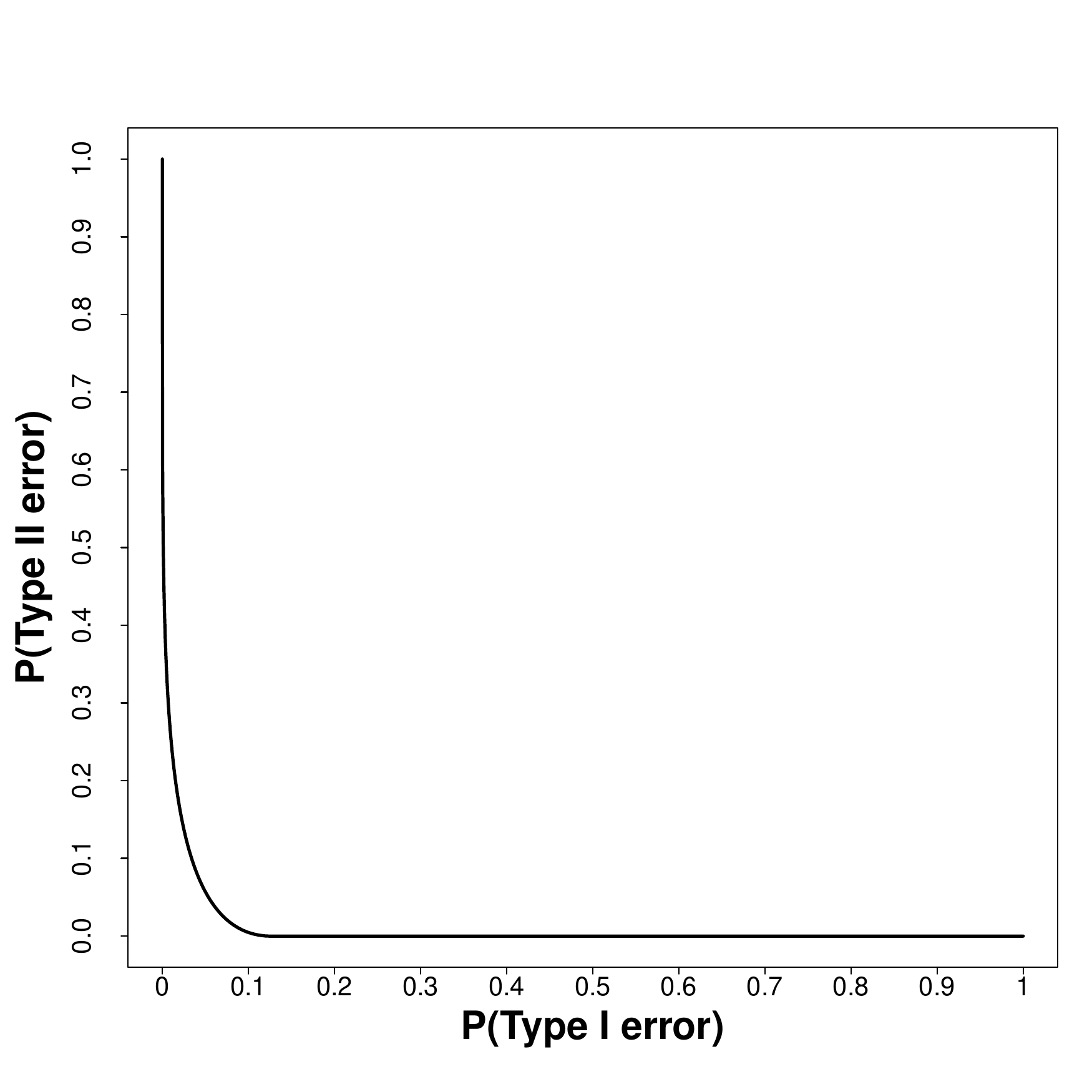}
  \caption{ROC curve for Type I and II error.}
  \label{fig: gfdr_roc1}
\end{subfigure}
\medskip
\caption{The solid line represents Type I error, and the dashed line represents Type II error as in Equations \eqref{typeI} and \eqref{typeII}. The error probabilities are calculated as a function of tail-area Fdr cutoff point $q$ with the rejection region in \eqref{R_comb1}. }
\label{fig: typeI_II}
\vspace{-0.2 cm}
\end{figure}

Figure \ref{fig: typeI_II} shows type I and type II errors as functions of cutoff points $q$ for the tail-area Fdr screening calculated from the simulated data following steps described in Section \ref{subs:power}. This may help in the choice of appropriate cutoff point $q$ for the main analysis.

In Appendix~\ref{app:Performance} we have included an overall performance comparison between the repeated sample splitting/screening procedure and the conventional Benjamini-Hochberg screening method over 100 different data set simulated from the Normal distribution setup used in this section. The comparison reveals that our approach, as a whole, performed better in terms of stability selection \cite{Meinshausen}. The sample splitting/screening approach could successfully detect all non-null genes, where the BH screening detected some non-null genes with high relative frequency but missed nearly half of them. In the results, a prominent specificity difference between the two approaches was apparent as well. Our approach almost never detected null genes as non-null, while the BH method erroneously picked up almost all null genes as non-null  multiple times over the course of 100 simulations (Appendix~\ref{app:Performance} contains further details).

\subsubsection{Null Model Simulation (Specificity Study)}
\label{subs:simulation2}
Next, we consider the null model, i.e., the scenario where both the experiment and the control group data come from the same population distribution. In this case, all hypotheses under study belong to the null class. Hence any classification or detection of a non-null case will essentially be a false discovery. To understand if a given procedure works well in this scenario, we can examine the percentage of time the procedure can assign a hypothesis correctly to the null class or the ``specificity" of the method.

\begin{figure}
\centering
\vspace{-3 cm}
\begin{subfigure}{.9\textwidth}
  \centering
  \includegraphics[width=.95\linewidth]{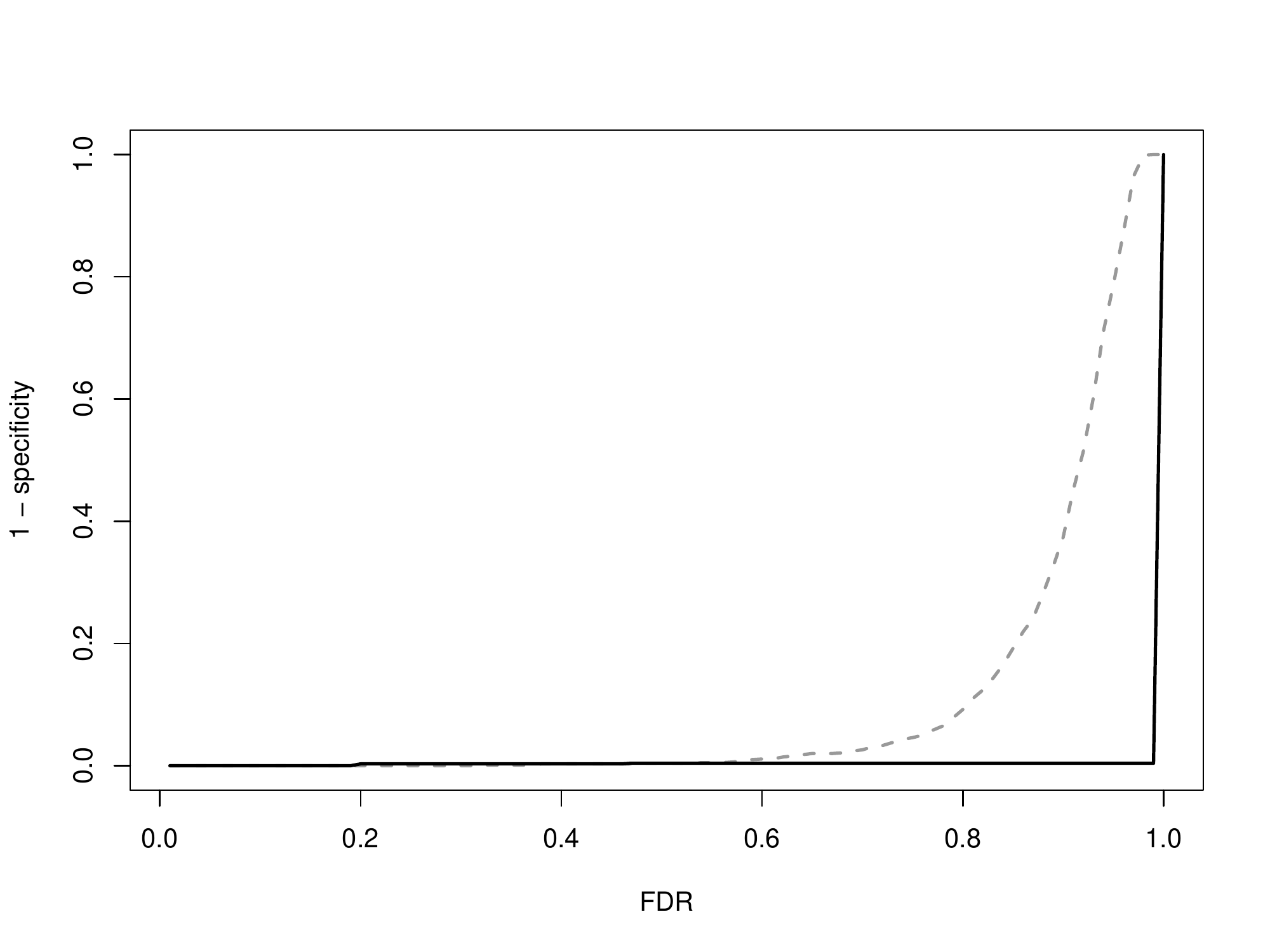}
  \vspace{-0.1 cm}
  \caption{Specificity comparison between the Benjamini-Hochberg method and the proposed method. The dark solid line represents the Benjamini-Hochberg FDR method, and the gray dashed line is for the proposed method.}
  \label{fig:null_specificity1}
\end{subfigure}%
\par
\vspace{-0.5 cm}
\begin{subfigure}{.9\textwidth}
  \centering
  \includegraphics[width=0.95\linewidth]{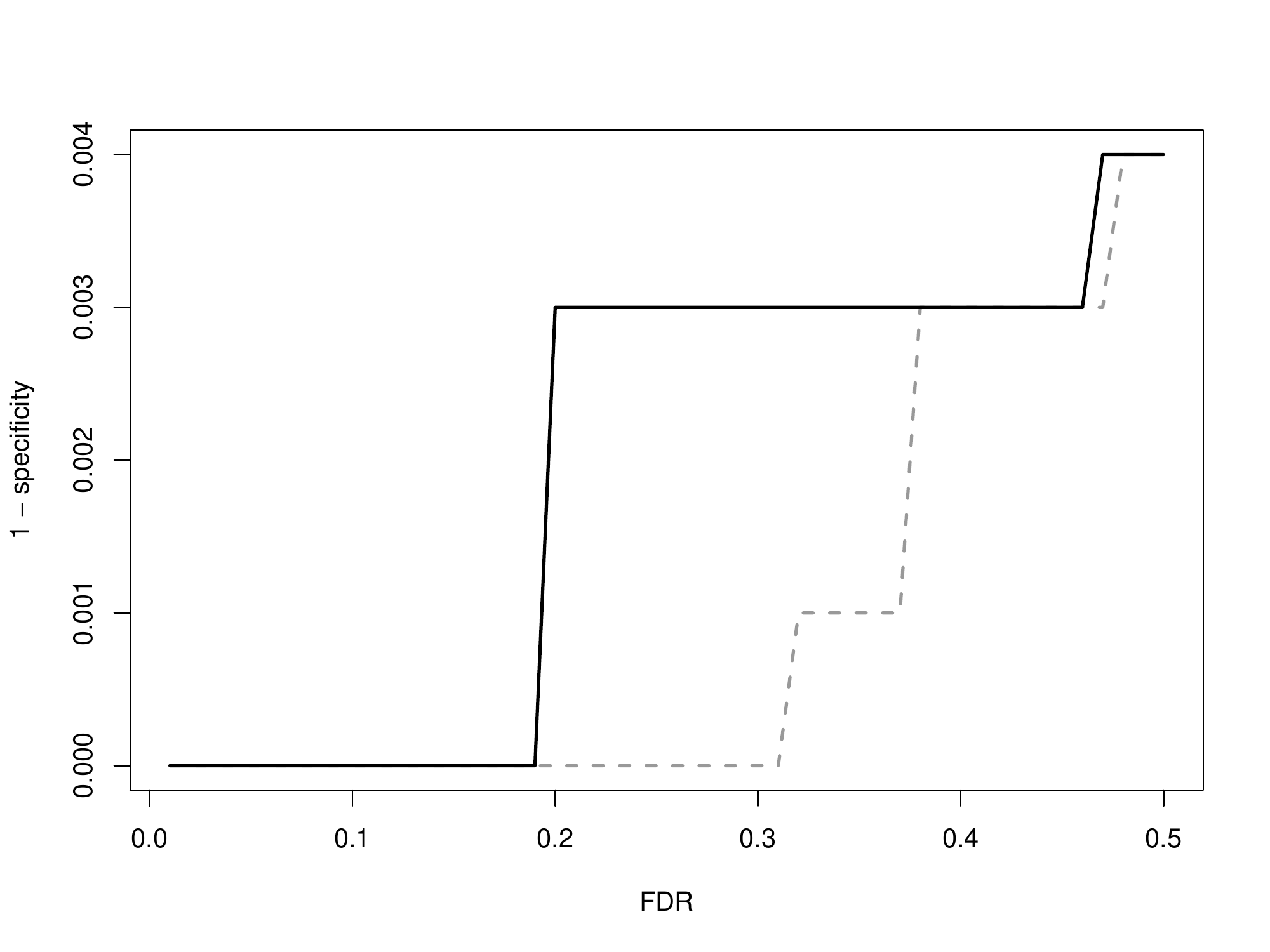}
  \vspace{-0.1 cm}
  \caption{The zoomed in version of  Figure \ref{fig:null_specificity1}. The dark solid line represents  the Benjamini-Hochberg FDR method, and the gray dashed line is for the proposed method. }
 \label{fig:null_specificity2}
\end{subfigure}
\bigskip
\caption{Specificity comparison plot for the proposed method with the BH method. The x-axis shows the FDR cutoff point used for screening. The y-axis represents $(1-specificity)$. Therefore  smaller hight of the plot indicates higher specificity, i.e., a  higher rate of correct classification of null cases.}
\label{fig:null_specificity}
\vspace{-2 cm}
\end{figure}

We simulated continuous data from 100 subjects for 1000 genes, from the normal distribution with the mean $\mu=6$, and standard deviation $\sigma=2$. The first 50 subjects and the next 50 subjects respectively constitute the control group and the treatment group, which obviously come from the same distribution. Figure~\ref{fig:null_specificity} illustrates the specificity comparison of the BH method (dark solid line) and the proposed method (gray dashed line). The $y$-axis presents $(1-specificity)$, so smaller hight of the plot indicates higher specificity, i.e., a higher rate of correct classification of null cases. The plot shows that the proposed method has slightly higher specificity for moderate FDR ($<0.5$), but lower specificity for large FDR ($>0.5$). Note that, for multiple testing screening factors with small FDR are considered to be true discoveries, and Figure~\ref{fig:null_specificity} shows that the proposed method works better in that crucial region.

\subsubsection{Discrete Data Simulation}
\label{subs:simulation3}
Here we consider a simulation of a discrete data for 100 subjects and 1000 genes, in which the first 50 subjects are assumed to form the control group, and the other 50 subjects form the treatment group. The control group values are simulated from the negative binomial distribution with the mean $\mu_0=1000$ and the dispersion $\phi=0.5$, in notation, $NB(\mu_0=1000,\phi=0.5)$ (we are using the Negative-Binomial parameterization used in \cite{robinson2007small}). The treatment group data is simulated from the mixture of negative binomial distributions,  $0.97NB(\mu_0=1000,\phi)+0.03NB(\mu_1=1500,\phi=0.5)$, where the first 30 genes are the non-null genes that are simulated from the negative binomial distribution with mean $\mu_1=1500$ and the same dispersion value $\phi=0.5$ as the control group.

\begin{table}[htp]
\centering
\caption{Screening results for simulated discrete data where freq and rfreq are calculated from the proposed method in 500 sample splits.  Genes 1 to 30 are set to be the non-null genes. The values in the column $\mathrm{FDR_{BH}}$  are calculated based on the BH method by using the whole data. The last column presents the p-values for each gene from the whole data likelihood ratio tests that compare the treatment group and the control group averages.}
\label{tab:rna_sim}
\medskip
\begin{tabular}{crrrr}
  \hline
Gene\_ID & freq & rfreq & $\mathrm{FDR_{BH}}$ & pvalue \\
  \hline
22 & 44 & 0.088 & 0.001 & 1.01E-06 \\
  27 & 17 & 0.034 & 0.043 & 2.55E-04 \\
  5 & 14 & 0.028 & 0.031 & 6.95E-05 \\
  25 & 11 & 0.022 & 0.031 & 9.36E-05 \\
  209 & 9 & 0.018 & 0.033 & 1.30E-04 \\
  21 & 9 & 0.018 & 0.081 & 9.37E-04 \\
  4 & 8 & 0.016 & 0.034 & 1.71E-04 \\
  2 & 7 & 0.014 & 0.048 & 3.34E-04 \\
  19 & 6 & 0.012 & 0.054 & 4.70E-04 \\
  193 & 5 & 0.010 & 0.081 & 9.68E-04 \\
  29 & 5 & 0.010 & 0.054 & 5.22E-04 \\
   \hline
\end{tabular}
\end{table}

We used 500 random splits to analyze the data. By using the Fdr threshold 0.05 and the critical detection frequency threshold 0.01, our method found 11 significant genes including 9 true discoveries and had 2 false discoveries. Alternatively, the conventional BH method detected 7 genes including 6 true discoveries and one false discovery by using the threshold 0.05.  Screening results are shown in Table~\ref{tab:rna_sim}, where the columns freq and rfreq illustrate the detection frequencies and detection relative frequencies of the significant genes. The values in the column $\mathrm{FDR_{BH}}$  are calculated based on the BH method by using the whole data. The last column presents the p-values for each gene from the whole data likelihood ratio tests that compare the treatment group and the control group averages. Table~\ref{tab:rna_sim} shows that genes 19, 21 and 29 were detected more than 5 times by our method, while had  $\mathrm{FDR_{BH}}$  greater than the threshold 0.05, hence would have been missed by the BH screening.

\subsubsection{Discrete Data Simulation with Larger Signal Proportion}
\label{subs:simulation4}
We repeated the discrete data simulation setup from Subsection \ref{subs:simulation3} but with 100 non-null genes in the treatment group instead of 30.  That is, data was simulated for 1000 genes for 50 control subjects and 50  treatment group subjects, with the control group data generated from a negative binomial distribution $NB(\mu_0=1000,\phi=0.5)$, and the treatment group data from a mixture of negative binomial distributions,  $0.9NB(\mu_0=1000,\phi)+0.1NB(\mu_1=1500,\phi=0.5)$, where the first 100 genes are set to be the non-null genes and  are simulated from the negative binomial distribution with mean $\mu_1=1500$.

The analysis was done with 500 random sample splits using the proposed method. By using the FDR threshold 0.05 and the relative significant frequency 0.02, our method detected 46 significant genes, including 39 true discoveries and 7 false discoveries. Comparatively, the BH method detected 39 genes, including 34 true discoveries and 5 false discoveries, respectively. The power of the proposed method was 0.39, and the power of the BH method was 0.34. Table~\ref{tab:large_signal} presents the screening results that includes the detection frequencies and relative frequencies of all 46 genes along with their corresponding BH method FDR. The table shows that the proposed method detected non-null genes 27, 47, 80, 85, and 18  at least 10 times, but had $\mathrm{FDR_{BH}}$ greater than 0.05, hence would have been missed by the BH method with the same 0.05 FDR threshold value.

\begin{center}
\begin{longtable}[ht]{crrrr}
\caption{Screening results for simulated discrete data with a larger proportion of the signal where genes 1 to 100 are set to be the non-null genes. The freq and rfreq are calculated from the proposed method in 500 sample splits.  The values in the column $\mathrm{FDR_{BH}}$  are calculated based on the BH method by using the whole data. The last column presents the p-values for each gene from the whole data likelihood ratio tests that compare the treatment group and the control group averages. The proposed method discovered 46 genes by using the FDR threshold 0.05, and the significant frequency threshold 0.02, respectively.}
\medskip
\label{tab:large_signal}
\\
\hline
Gene\_ID & freq & rfreq & $\mathrm{FDR_{BH}}$ & pvalue \\
\hline
\endfirsthead

\multicolumn{5}{c}
{{\bfseries \tablename \ \thetable{} -- continued from previous page}}\\
\hline
Gene\_ID & freq & rfreq & $\mathrm{FDR_{BH}}$ & pvalue \\
\hline
\endhead

\hline
\multicolumn{5}{r}{continued on next page}\\
\hline
\endfoot

\hline
\endlastfoot
86 & 123 & 0.246 & 0.002 & 5.03E-06 \\
  58 & 121 & 0.242 & 0.002 & 9.67E-06 \\
  70 & 107 & 0.214 & 0.002 & 1.15E-05 \\
  56 & 101 & 0.202 & 0.003 & 2.47E-05 \\
  66 & 96 & 0.192 & 0.003 & 1.89E-05 \\
  94 & 96 & 0.192 & 0.002 & 1.32E-05 \\
  55 & 93 & 0.186 & 0.002 & 1.28E-05 \\
  92 & 93 & 0.186 & 0.002 & 8.33E-06 \\
  21 & 77 & 0.154 & 0.006 & 8.97E-05 \\
  24 & 69 & 0.138 & 0.004 & 3.61E-05 \\
  25 & 66 & 0.132 & 0.003 & 2.41E-05 \\
  209 & 63 & 0.126 & 0.004 & 4.74E-05 \\
  23 & 53 & 0.106 & 0.012 & 1.86E-04 \\
  77 & 51 & 0.102 & 0.009 & 1.42E-04 \\
  72 & 50 & 0.100 & 0.004 & 5.31E-05 \\
  91 & 49 & 0.098 & 0.060 & 2.59E-03 \\
  83 & 47 & 0.094 & 0.024 & 7.30E-04 \\
  39 & 46 & 0.092 & 0.017 & 4.03E-04 \\
  82 & 45 & 0.090 & 0.005 & 6.10E-05 \\
  32 & 42 & 0.084 & 0.018 & 4.66E-04 \\
  99 & 42 & 0.084 & 0.017 & 3.95E-04 \\
  12 & 37 & 0.074 & 0.031 & 1.15E-03 \\
  57 & 34 & 0.068 & 0.013 & 2.64E-04 \\
  74 & 34 & 0.068 & 0.013 & 2.79E-04 \\
  8 & 29 & 0.058 & 0.013 & 2.52E-04 \\
  95 & 29 & 0.058 & 0.012 & 1.96E-04 \\
  35 & 28 & 0.056 & 0.018 & 5.00E-04 \\
  73 & 27 & 0.054 & 0.017 & 3.92E-04 \\
  97 & 25 & 0.050 & 0.031 & 1.07E-03 \\
  29 & 22 & 0.044 & 0.018 & 4.80E-04 \\
  22 & 21 & 0.042 & 0.013 & 2.80E-04 \\
  68 & 20 & 0.040 & 0.029 & 8.96E-04 \\
  193 & 18 & 0.036 & 0.018 & 4.46E-04 \\
  42 & 18 & 0.036 & 0.031 & 1.10E-03 \\
  60 & 18 & 0.036 & 0.024 & 7.26E-04 \\
  27 & 17 & 0.034 & 0.059 & 2.46E-03 \\
  927 & 15 & 0.030 & 0.029 & 9.42E-04 \\
  181 & 14 & 0.028 & 0.106 & 6.23E-03 \\
  763 & 14 & 0.028 & 0.031 & 1.22E-03 \\
  4 & 13 & 0.026 & 0.031 & 1.17E-03 \\
  117 & 12 & 0.024 & 0.031 & 1.05E-03 \\
  706 & 12 & 0.024 & 0.064 & 2.82E-03 \\
  47 & 11 & 0.022 & 0.090 & 4.53E-03 \\
  80 & 11 & 0.022 & 0.057 & 2.27E-03 \\
  85 & 11 & 0.022 & 0.077 & 3.75E-03 \\
  18 & 10 & 0.020 & 0.057 & 2.35E-03 \\
   \hline
\end{longtable}
%\vspace{-1 cm}
\end{center}

%%%%%%%%%%%%%%%%%%%%%%%%%%%%%
%%%%%%%%%%%%%%%%%%%%%%%%%%%%
%\newpage
\section{Discussion}
\label{s_discussion}
\textbf{\underline{Sample Splitting:}}

 In a multiple testing situation, if the entire available data is used for the fitting of a null/non-null mixture model, then using the same model for the non-null detection again from that data may affect the screening process due to overfitting of the empirical model. The sample splitting in the proposed method allows a part of the available information to be used for model building, and the other part for screening significant cases hence avoids that drawback. Further, when the data is randomly split multiple times, it produces a different (may not be disjoint) modeling set each time. When models are fitted based on these different splits, it helps to neutralize the effect of sources of variation (noise) other than the one that is of interest in a study.

The use of only partial information for the model building part may lead to some loss of power for an individual modeling split, but repeated sample splitting and combining the resulting rejection regions overcomes that. In fact, in all simulation studies, the proposed method had a higher number of true discoveries than the existing whole data fit method. In models with higher signal proportion, this difference was more pronounced. However, the combined rejection region also accumulates false discoveries and reduces precision, but the use of a critical detection frequency threshold is designed to attain any fixed precision level for the entire process.

\textbf{\underline{F-association Plots:}}

 The other benefit of  the repeated sample splitting is the F-association plots that are constructed based on the detection frequencies across the repeated splits.  The simulation study shows that if some screened non-null cases are indeed associated, that relation is captured  in the group structure of the F-association plot (see Appendix \ref{app:F_association} for a heuristic justification). Here we want to emphasize that the F-association plots generated by this method should not be used as a ``proof" of group behavior among the cases. If two unrelated non-null cases deviate strongly from the null distribution, they are bound to be frequently detected as significant, no matter how the data is split. In that case, these two unrelated non-null cases may appear in the detected sets repeatedly at the same time and consequently will show up in the F-association plot as a group. Therefore, the F-association plot is not intended to use as a basis for a causal relation.

Thus we recommend that these plots be used as an exploratory tool that precedes further investigation to establish possible causal relationship between cases that show high concurrent detection frequencies. When a study includes thousands of cases, at least some starting point for an exploratory network analysis can be highly useful and cost-effective.

The relevance and effectiveness of the proposed method can be explained particularly well in a study where the main goal is the identification of groups of differentially expressed genes. Such studies are commonly used to identify the genes that are associated with specific biological behavior. Genes detected through the proposed methodology can be selected by experimental biologists for detailed follow-up functional studies. For example, a biologist may look into the few most frequently identified significant genes to distinguish the ``regulators". A systematic gene knockout experiment, conducted on the sets that  appear in the F-association plot as a group, can reveal the effect of individual genes on the biological response of interest. This can provide a novel starting point for the elucidation of gene networks or hierarchical regulation patterns in a biological system. Thus, the proposed analysis can guide an exploratory biological study where, instead of experimental investigations of the effect of every gene, a small subset of significant ones can be selected for further experimentation to establish their individual or collective role in the biological response of interest.

\textbf{\underline{Conditional Independence:}}

An interesting question can be posed: ``When does the empirical Bayes model \eqref{eq:main_model} work?" It of course works when the observations are i.i.d according to \eqref{eq:main_model}. But in many cases (in particular, the microarray example considered here), it is not correct. It does work, though, in pooling observations where the observations are conditionally independent. For example, in a microarray, groups of genes may be acting together but still conditionally independent. This is a typical argument used in multiple hypothesis testing cases \cite{Karlin81, Sarkar97, Huque16}. See Appendix \ref{app:Pooling data} for an explanation of why the empirical Bayes approach works in this case.

In conclusion, we present  a  method that can be used for identifying significant cases when carrying out a  large number of simultaneous tests. We propose a cross-validation type analysis where a part of the available information goes into the understanding of the underlying process or model fitting, while the other part goes into screening for extreme cases. Random splitting and repeated screening provide a way to reduce the noise (other sources of variation) in the analysis. We present a way of controlling the overall chance of false discoveries that arise from combining the screening results from all splits. As a by-product we get an exploratory look into  the associative pattern between significant cases.
%%%%%%%%%%%%%%%%%%%%%%%%%%%%
\section{Software} The analysis were done using $\texttt{R}$. The relevant codes are available upon request. An $\texttt{R}$ package for the proposed analysis method is available  for general use. The R package $\texttt{tiltmod}$ can be downloaded via: \url{https://github.com/chongma1989/tiltmod}.

\bibliographystyle{authordate1}
\bibliography{CSDApaper_bib}

\begin{thebibliography}{}

\bibitem[\protect\citename{Allison {\em et~al.}, }2002]{Allison}
Allison, D, Gadbury, G.~L, Moonseong, H, Fernandez, J.R, Lee, C.K, Prolla, T.A,
  \& Weindruch, R. 2002.
\newblock A mixture model approach for the analysis of microarray gene
  expression data.
\newblock {\em Computational Statistics and Data Analysis}, {\bf 39}, 1--20.

\bibitem[\protect\citename{Anders \& Huber, }2010]{anders2010diff}
Anders, S, \& Huber, W. 2010.
\newblock Differential expression analysis for sequence count data.
\newblock {\em Genome Biology}, {\bf 11}, R106.

\bibitem[\protect\citename{Bailey {\em et~al.}, }2016]{Baily}
Bailey, D.H, Borwein, J.M, \& Stodden, V. 2016.
\newblock Facilitating reproducibility in scientific computing: Principles and
  practice.
\newblock {\em Pages  205 -- 232 of:} Atmanspacher, Harold, \& Maasen, Sabine
  (eds), {\em Reproducibility: Principles, Problems, Practices and Prospect}.
\newblock New York: John Wiley and Sons.

\bibitem[\protect\citename{Benjamini \& Hochberg, }1995]{Benjamini}
Benjamini, Y, \& Hochberg, Y. 1995.
\newblock Controlling the false discovery rate: A practical and powerful
  approach to multiple testing.
\newblock {\em Journal of Royal Statistical Society: Series B}, {\bf 57}(1),
  289--300.

\bibitem[\protect\citename{Berrar, }2019]{BERRAR}
Berrar, D. 2019.
\newblock Cross-Validation.
\newblock {\em Pages  542 -- 545 of:} Ranganathan, Shoba, Gribskov, Michael,
  Nakai, Kenta, \& Schönbach, Christian (eds), {\em Encyclopedia of
  Bioinformatics and Computational Biology}.
\newblock Oxford: Academic Press.

\bibitem[\protect\citename{Dickhaus, }2014]{Dickhaus}
Dickhaus, T. 2014.
\newblock {\em Simulteneous Statistical Inference, with the Application in the
  Life Sciences}.
\newblock Springer-Verlag.

\bibitem[\protect\citename{Efron, }2007]{Efron07}
Efron, B. 2007.
\newblock Size, power and false discovery rates.
\newblock {\em Annals of Statistics}, {\bf 35}(4), 1351--1377.

\bibitem[\protect\citename{Efron, }2008]{Efron08}
Efron, B. 2008.
\newblock Microarrays, empirical {Bayes}, and the two-groups model.
\newblock {\em Statistical Science}, {\bf 23}, 1--22.

\bibitem[\protect\citename{Efron, }2010]{Efron10}
Efron, B. 2010.
\newblock {\em Large-Scale Inference: Empirical Bayes Methods for Estimation}.
\newblock Cambridge, UK: Cambridge University Press.

\bibitem[\protect\citename{Efron \& Morris, }1973]{Morris}
Efron, B, \& Morris, C. 1973.
\newblock Stein's estimation rule and its competitors--an empirical {Bayes}
  approach.
\newblock {\em Journal of the American Statistical Association}, {\bf 68}(341),
  117--130.

\bibitem[\protect\citename{Grego {\em et~al.}, }1990]{Grego}
Grego, J, Hsi, H-L, \& Lynch, J. 1990.
\newblock A Strategy for analyzing mixed and pooled exponentials.
\newblock {\em Applied Stochastic Models And Data Analysis.}, {\bf VI}, 59--70.

\bibitem[\protect\citename{Harell {\em et~al.}, }1996]{Harell}
Harell, FE~Jr1, Lee, KL, \& Mark, DB. 1996.
\newblock Multivariable prognostic models: issues in developing models,
  evaluating assumptions and adequacy, and measuring and reducing errors.
\newblock {\em Statistics in Medicine}, {\bf 15}(4), 361--87.

\bibitem[\protect\citename{Hirakawa {\em et~al.}, }2007]{hirakawa}
Hirakawa, A, Sato, Y, Sozu, T, Hamada, C, \& Yoshimura, I. 2007.
\newblock Estimating the False Discovery Rate Using Mixed {N}ormal Distribution
  for Identifying Differentially Expressed Genes in Microarray Data Analysis.
\newblock {\em Cancer Informatics}, {\bf 3}, 140--148.

\bibitem[\protect\citename{Huque, }2016]{Huque16}
Huque, MF. 2016.
\newblock Validity of the {Hochberg} procedure revisited for clinical trial
  applications.
\newblock {\em Statistics in Medicine}, {\bf 35}(1), 5--20.

\bibitem[\protect\citename{Karlin \& Rinott, }1981]{Karlin81}
Karlin, S, \& Rinott, Y. 1981.
\newblock Total positivity properties of the absolute value multinormal
  variables with applications to confidence interval estimates and related
  probabilistic inequalities.
\newblock {\em Annals of Statistics}, {\bf 9}, 1035--1049.

\bibitem[\protect\citename{Mathur {\em et~al.}, }2011]{Mathur}
Mathur, R, Schaffer, J, Land, W, J~Heine, J, Hernandez, J, \& Yeatman, T. 2011.
\newblock Perturbation and candidate analysis to combat overfitting of gene
  expression microarray data.
\newblock {\em International journal of computational biology and drug design},
  {\bf 4}, 307--3015.

\bibitem[\protect\citename{Meinshausen \& Buhlmann, }2010]{Meinshausen}
Meinshausen, N, \& Buhlmann, P. 2010.
\newblock Stability Selection.
\newblock {\em Journal of Royal Statistical Society: Series B}, {\bf 72}(4),
  417--473.

\bibitem[\protect\citename{Muralidharan, }2010]{Murlidharan}
Muralidharan, O. 2010.
\newblock An Empirical {Bayes} Mixture Method for Effect Size and False
  Discovery Rate.
\newblock {\em Annals of Applied Statistics}, {\bf 4}(1), 422--438.

\bibitem[\protect\citename{Muralidharan {\em et~al.},
  }2012]{muralidharan2012detecting}
Muralidharan, O, Natsoulis, G, \& Bell, J et~al. 2012.
\newblock Detecting mutations in mixed sample sequencing data using empirical
  Bayes.
\newblock {\em The Annals of Applied Statistics}, {\bf 6}(3), 1047--1067.

\bibitem[\protect\citename{Pickrell {\em et~al.},
  }2010]{pickrell2010understanding}
Pickrell, JK, Marioni, JC, \& Pai, A et~al. 2010.
\newblock Understanding mechanisms underlying human gene expression variation
  with RNA sequencing.
\newblock {\em Nature}, {\bf 464}(7289), 768--772.

\bibitem[\protect\citename{Powers, }2011]{Powers}
Powers, DMW. 2011.
\newblock Evaluation: From Precision, Recall and {F}-Measure to {ROC},
  Informedness, Markedness \& Correlation.
\newblock {\em Journal of Machine Learning Technologies}, {\bf 2}(1), 37--63.

\bibitem[\protect\citename{{Raeder} {\em et~al.}, }2010]{Raeder}
{Raeder}, T, {Hoens}, T.~R, \& {Chawla}, N.~V. 2010 (Dec).
\newblock Consequences of Variability in Classifier Performance Estimates.
\newblock {\em Pages  421--430 of:} {\em 2010 IEEE International Conference on
  Data Mining}.

\bibitem[\protect\citename{Robbins, }1956]{Robbins}
Robbins, H. 1956.
\newblock An empirical {Bayes} approach to statistics.
\newblock {\em Proceedings of the Third Berkeley Symposium on Mathematical
  Statistics and Probability}, {\bf I}, 157--163.

\bibitem[\protect\citename{Robinson \& Smyth, }2007]{robinson2007small}
Robinson, MD, \& Smyth, GK. 2007.
\newblock Small-sample estimation of negative binomial dispersion, with
  applications to SAGE data.
\newblock {\em Biostatistics}, {\bf 9}(2), 321--332.

\bibitem[\protect\citename{Sarkar \& Chang, }1997]{Sarkar97}
Sarkar, SK, \& Chang, CK. 1997.
\newblock Simes' method for multiple hypotheses testing with positively
  dependent test statistics.
\newblock {\em Journal of the American Statistical Association}, {\bf 92},
  1601--1608.

\bibitem[\protect\citename{Simon {\em et~al.}, }2003]{Simon}
Simon, R, Radmacher, M, Dobbin, K, \& McShane, LM. 2003.
\newblock Pitfalls in the Use of DNA Microarray Data for Diagnostic and
  Prognostic Classification.
\newblock {\em Journal of the National Cancer Institute}, {\bf 95}(1), 14–18.

\bibitem[\protect\citename{Singh {\em et~al.}, }2002]{Singh}
Singh, D, Febbo, P, \& Ross, K et~al. 2002.
\newblock Gene expression correlates of clinical prostate cancer behavior.
\newblock {\em Cancer Cell.}, {\bf 1}, 203--209.

\bibitem[\protect\citename{Smyth, }2004]{smyth2004linear}
Smyth, GK. 2004.
\newblock Linear models and empirical Bayes methods for assessing differential
  expression in microarray experiments.
\newblock {\em Statistical applications in Genetics and Molecular Biology},
  {\bf 3}, Article 3.

\bibitem[\protect\citename{Subramanian \& Simon, }2013]{Subramanian}
Subramanian, J, \& Simon, R. 2013.
\newblock Overfitting in prediction models- is it a problem only in high
  dimension?
\newblock {\em Contemporary Clinical Trials}, {\bf 36}, 636--641.

\end{thebibliography}
\clearpage
%%%%%%%%%%%%%%%%%%%%%%%%%%%%%
%%%%%%%%%%%%%%%%%%%%%%%%%%%%%%%%%%%%%%%%%%%%%%%%%%%%%%

\appendix
\begin{center}
{\bf Appendix}
\end{center}

\section{F-association plots justification.}
\label{app:F_association}

When we are trying to detect genes that have significantly different levels of expression in the treatment group compared to that of the control group, the detection can be based on the difference in the average expression level between the two groups.

Let $D_i$ denote the average expression difference between the experiment and the control group for the $i^{th}$ gene. Genes that behave similarly in both groups should have $D_i$'s close to $0$, whereas genes that are truly different in these two groups should have $D_i$'s removed from $0$ in either a positive or a negative direction. These extreme genes will have smaller chances of being false discoveries. Thus an extreme gene detection criterion of the form $Fdr<q$ for some $q \in (0,1)$ is equivalent of $|D_i|>c$ for some $c>0$.

Let us consider two genes $G_1$ and $G_2$ that are truly different on average in the treatment group compared to the control group. Suppose $G_1$ is detected, that is, the observed value $d_1$ of $D_1$, or the average expression difference for $G_1$ is beyond the threshold; without loss of generality suppose $d_1>c$. We present an argument below that the chance that $G_2$ also will be detected is higher when $G_1$ and $G_2$ have some level of dependency as compared to the scenario when $G_1$ and $G_2$ behave completely independently.

Similar to the simulation study in Section \ref{subs:simulation}, we assume the $D_i$'s follow a Normal distribution (not an entirely hypothetical scenario if we have large enough sample sizes in the experiment and the control groups). If $G_1$ and $G_2$ function independently, we assume $D_1, D_2$ are $i.i.d$ $N(0,1)$ variables. If $G_1$ and $G_2$ are not independent we assume $\{D_1, D_2\}\sim N(\bm{\mu}, \bm{\Sigma})$. with $\bm{\mu}=\left[
                \begin{array}{c}
                  0\\
                 0\\
                \end{array}
              \right]$
and
$\bm{\Sigma}=\left[
                \begin{array}{cc}
                  1& \rho\\
                  \rho & 1\\
                \end{array}
              \right]
$.\\
(We will denote the left and right tail probabilities of the standard Normal distribution by $\Phi(\cdot)$ and $\bar{\Phi}(\cdot)$ respectively).

If the two genes behave independently and $G_1$ is detected, the chance that $G_2$ will also be detected is $P(D2>c|D_1=d_1)=P(D_2>c)=\bar{\Phi}(c)$ (here $\bar{\Phi}(c)$ is the right-tail area from $c$ in a standard Normal distribution). If the two genes are not independent then we have the following two possibilities:
  \begin{itemize}
  \item If $\rho>0$ and $G_1$ is detected, the chance that $G_2$ will also be detected is
   \[P(D2>c|D_1=d_1)=\bar{\Phi}\left(\frac{c-\rho d_1}{\sqrt{1-\rho^2}}\right)\]
  \item If $\rho<0$  and $G_1$ is detected,  the chance that $G_2$ will also be detected is
   \[P(D2<-c|D_1=d_1)=\Phi\left(\frac{-c-\rho d_1}{\sqrt{1-\rho^2}}\right)\]
\end{itemize}
If $\rho>0$  we have:
\begin{align*}
d_1>c\Rightarrow c-\rho d_1<c-\rho c \Rightarrow& \frac{c-\rho d_1}{\sqrt{1-\rho^2}}<\sqrt{\frac{1-\rho}{1+\rho}}\ c<c\\
\Rightarrow& \bar{\Phi}\left(\frac{c-\rho d_1}{\sqrt{1-\rho^2}}\right)>\bar{\Phi}(c)
\end{align*}
If $\rho<0$  we have:
\begin{align*}
d_1>c\Rightarrow -c-\rho d_1>-c-\rho c\Rightarrow& \frac{-c-\rho d_1}{\sqrt{1-\rho^2}}>-\sqrt{\frac{1+\rho}{1-\rho}}\ c>-c\\
\Rightarrow& \Phi\left(\frac{-c-\rho d_1}{\sqrt{1-\rho^2}}\right)>\Phi(-c)=\bar{\Phi}(c).
\end{align*}
Thus in both cases, the chance that gene $G_2$ is detected when $G_1$ is detected increases if the two genes are not independent. Although we have presented the idea using the Normal distribution, the argument  can be used for any location-scale family with proper parameter quantifying the dependence.

In the F-association plot, the detection (relative) frequencies are the empirical estimates of the detection probabilities. Thus if two genes are detected together frequently, there is a chance that they are not functioning independently. However, if  two genes  are independent but truly very different in the treatment group versus the control group, both of them will be detected almost always and will be  a part of the screened set for many splits. Thus the high concurrent detection frequency is not sufficient argument for claiming two genes are related; it should rather be considered as a necessary condition that can serve as the starting point for biological experiments for exploring gene association networks.

\section{Pooling data}
\label{app:Pooling data}
In situations where we have clusters of observations and within a cluster the observations are conditionally independent, pooling of them can result in the observations being i.i.d. from the pooled/mixed distribution model. \cite{Grego} was one of the first to suggest the use of mixed distribution methodology to analyze such data when the observations are exponentially distributed. Here we provide a justification of this type of analysis for more complicated situations such as the one considered in this paper.

To see this, consider $k$ clusters, where there are $n_i$ observations, $X_{i,1},\ldots, X_{i,n_i}$ in cluster $C_i$.  Consider the situation where the joint density of all the observations can be written as
\begin{equation}
\label{eq:pooled1}
g(I)\prod\limits_{j=1}^k\left[\prod\limits_{m=1}^{n_j}f\left(\left.x_{j,m}\right|\lambda_{j,m}\right)g_{j,m}\left(\left.\lambda_{j,m}\right|I_j\right)\right]
\end{equation}
where $\bm{I}=(I_1,\ldots,I_k)$ is a vector of indicator variables indicating if the clusters are in the background state $(I_j=0)$ or in the signal state $(I_j=1)$. Note that, if $g(\bm{I})=\prod\limits_{j=1}^kg(I_j)$ (the indicator variables are independent), then the $X'$s are independent.

For example, a cluster might be a biological network of genes where the indicator $I=0$ denotes that the genes in the network are not being differentially expressed and any expressed genes are simply background while if $I=1$ the network is being differentially expressed (signal).

Typically, we do not know the clusters/networks and are simply pooling the data.  Thus, from \eqref{eq:pooled1}
\begin{equation}
\label{eq:pooled2}
\left\{(X_{j,m},\Lambda_{j,m})\right\} \mbox{ given } I \mbox{ are independent.}
\end{equation}
Notice that this is almost an empirical Bayes or a mixture model formulation except that the distributions of the observations are not identically distributed.\\

However, notice that, given $(\bm{I},\bm{\Lambda})$, the conditional distribution of the $X$'s is given by  $X_{j,m}|(\bm{I},\bm{\Lambda})\sim f(x_{j,m}|\Lambda_{j,m})$.  Thus, if we pool the data, then, given $(\bm{I},\bm{\Lambda})$, the resulting $X$'s have marginal mixed density,
\begin{equation}
f(x)=\int f(x|\lambda)m(d\lambda)
\label{eq:pooled3}
\end{equation}

with support $\bm{\Lambda}$ where the point masses are determined by $\{g_{j,m}\left(\left.\Lambda_{j,m}\right|I_j\right)\}$.
That is, we are observing $X$'s for each gene from marginal density \eqref{eq:pooled3}, where they can be considered conditionally independent in the pooled data. Thus, given $(\bm{I},\bm{\Lambda})$, and \eqref{eq:pooled2} the form of \eqref{eq:pooled3} justifies the  use of the mixture distribution/empirical Bayes that we developed.

\section{Performance comparison}
\label{app:Performance}
To compare the general performance of the proposed method with the conventional Benjamini- Hochberg method, we used the simulation setup from Subsection~\ref{subs:simulation1} with 1000 variables where variables 1 to 30 are set to be non-null, and variables 31 to 1000 are set to be null variables.

\begin{figure}[!h]
\centering
%\vspace{-3 cm}
\begin{subfigure}{1\textwidth}
  \centering
  \includegraphics[width=.97\linewidth]{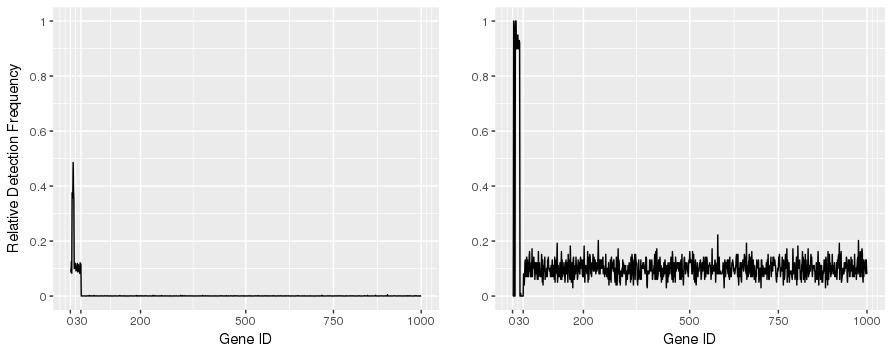}
  \vspace{0.1 cm}
  \caption{Performance comparison between the Benjamini-Hochberg method and the proposed method. The left panel presents the chance of getting detected as significant by the proposed method, and the right panel shows the chance of getting detected by the Benjamini-Hochberg method.}
  \label{fig:perf1}
\end{subfigure}%
\par
\vspace{0.5 cm}
\begin{subfigure}{1\textwidth}
  \centering
  \includegraphics[width=0.97\linewidth]{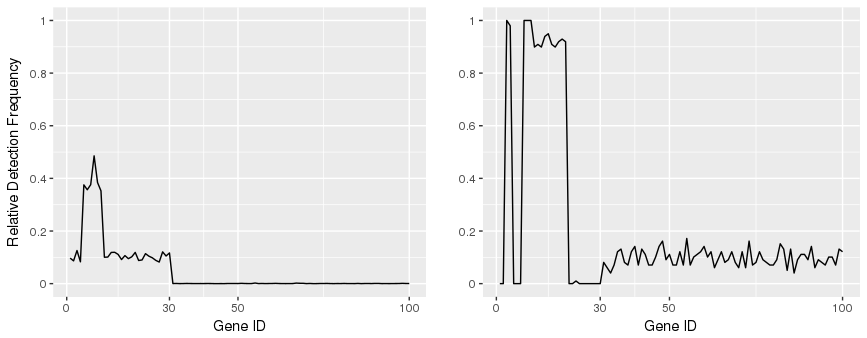}
  \vspace{0.1 cm}
  \caption{The zoomed in version of  Figure \ref{fig:perf1}. The left panel shows the relative detection frequency by the proposed method, and the right panel shows the same by the BH method.}
 \label{fig:perf2}
\end{subfigure}
\bigskip
\caption{Variables 1 to 30 are true non-null variables, and 31 to 1000 are null variables. The y-axis indicates the relative frequency of detection as non-null for each variable (Gene). The proposed method in left panels shows the relative frequency of detection for every 970 null variables is almost zero, and the relative frequency of detection for every 30 non-null variables is significantly larger than zero over 100 different simulations from Table~\ref{table_sim_par} setup. The BH method, shown in the right panels, had high relative frequency of detection for some non-null variables; however, it missed quite a few of them whereas almost all null variables were falsely detected by the BH method (non-zero relative detection frequency) in some of the 100 simulations.}
\label{fig:perform}
%\vspace{-2 cm}
\end{figure}

We repeated the simulation 100 times. For each simulated data set, we applied 500 cross-verifications (splitting/screening) following the proposed method where a variable was detected significant if tail-area Fdr$<0.1$, and obtained the relative frequency of detection in 500 screenings for each variable. The average relative detection frequency from all 100 different simulations was used for performance comparison in Figure \ref{fig:perform}. For each variable, this average relative frequency estimates the probability of getting screened as non-null by the method.

We also applied the Benjamini-Hochberg screening for each of the simulation data sets. In this case, a variable was detected as significant if FDR$_{\rm{BH}}<0.1$. Here, for each variable,  the relative frequency of detection out of 100 simulations can be used to estimate the probability of getting screened as non-null by the BH method.

Figure~\ref{fig:perform} compares the relative frequency of detection from the proposed method and the BH method over all 100 simulations. The left panels in the figure represent results from the sample splitting/screening analysis, and the right panels show the same from the BH method.  It is evident from the plot that the proposed method performs better in terms of stability selection \cite{Meinshausen} than the conventional FDR approach. Recall that in the simulation, the first 30 variables are set to be non-null, and the next 970 variables are null. Figure~\ref{fig:perform} shows that our method could distinguish the null and non-null variables better. The left panels in Figure~\ref{fig:perform} show that the proposed method was able to detect every true non-null variable as significant with nonzero probability.  Whereas, since the conventional FDR approach merely utilizes the data once, it is very likely to yield biased selections. While the conventional FDR (BH) approach detected some of the non-null variables with very high relative frequency, it failed to select close to half of the 30 non-null genes correctly as significant.

 The specificity difference between the two modes of screening is especially pronounced in Figure~\ref{fig:perform}. The repeated sample splitting/screening almost never selected null genes 31 to 1000 as non-nulls (relative frequency of detection almost 0 in left panels of Figure~\ref{fig:perform}). In comparison, the BH method picked all 970 null genes erroneously as non-nulls in several simulations (relative frequency of detection is non-zero for null genes 31 to 1000).
\end{document}